\documentclass[12pt,a4paper,hypertex]{article}
\usepackage{jheppub}

\allowdisplaybreaks[1]
\usepackage{bm}
\usepackage{subfigure}

\title{The fate of chaotic strings in a confining geometry}

\author[1]{Takaaki Ishii,}
\author[2]{Keiju Murata}
\author[3]{and Kentaroh Yoshida}
\affiliation[1]{University of Colorado, 390 UCB, Boulder, CO 80309, USA}
\affiliation[2]{Keio University, 4-1-1 Hiyoshi, Yokohama 223-8521, Japan}
\affiliation[3]{Department of Physics, Kyoto University, Kitashirakawa Oiwake-cho, Kyoto 606-8502, Japan}
\emailAdd{takaaki.ishii@colorado.edu}
\emailAdd{keiju@phys-h.keio.ac.jp}
\emailAdd{kyoshida@gauge.scphys.kyoto-u.ac.jp}

\abstract{%
We study chaotic motion of classical closed strings in the five-dimensional Anti-de Sitter (AdS) soliton spacetime. We first revisit classical chaos using a cohomogeneity-1 string ansatz. We then consider turbulent behaviors of the classical strings when the spatial dependence of the string world-sheet is included. Sensitivity to initial conditions in chaotic systems suggests that the string under chaos tends to stretch in the AdS soliton spacetime in a Lyapunov timescale. In this process, the orbital angular momentum transfers to internal spin due to the turbulence on the string. It follows that the string stays around the tip of the AdS soliton with a jumbled profile. We evaluate the spectra of conserved quantities 
and discuss their universal power-law scalings in the turbulent behaviors. 
}

\preprint{KUNS-2645}


\begin{document}

\maketitle

\section{Introduction}

The gauge/string duality is one of the most fascinating subjects in string theory.
A typical example is the conjectured equivalence between type IIB superstring theory on the AdS$_5\times S^5$ spacetime and $\mathcal{N}=4$ $SU(N)$ 
4D super Yang-Mills theory at large $N$~\cite{M,GKP,W}. 
Generalization has also been considered to the dualities between string theories in asymptotically AdS spacetimes and their dual field theories. 
The gauge/string duality brings a strong motivation and importance to understand dynamics of fundamental strings in asymptotically AdS spacetimes.

Nowadays, it has been well recognized that an integrable structure exists 
behind the AdS/CFT correspondence (for a big review, see \cite{review}). 
In particular, the supercoset structure of AdS$_5\times S^5$ is really crucial 
for showing the classical integrability in the string-theory side \cite{BPR}. 
The integrability enables us to use powerful techniques in studying the conjectured relations non-perturbatively. 
An enormous number of studies have been done along this direction.

This nice property is, however, exceptional, and in general the dualities between 
string theories and gauge theories do not enjoy the integrability. 
Most theories are not integrable, and they exhibit chaos.
For example, it is known that classical strings in AdS soliton spacetime~\cite{Horowitz:1998ha}, which is given just by an one-parameter deformation of the pure AdS spacetime, admit the chaos~\cite{Basu:2011dg}.
One can also find chaotic strings in various background geometries including AdS$_5\times T^{1,1}$ spacetime\footnote{The $T^{1,1}$ is a 5D Sasaki-Einstein manifold \cite{Candelas} 
and the AdS$_5\times T^{1,1}$ background is dual to a 4D superconformal field theory \cite{KW}. 
The coset construction of $T^{1,1}$ has been refined in \cite{CMY}.}
and black hole backgrounds \cite{T11,T11-ppwave,BH,WQCD,D-brane,complex-beta,NR,gamma}.\footnote{Very recently, by applying Melnikov's method \cite{Melnikov1,Melnikov2},  
it has been shown in \cite{AKY} that chaotic string solutions exist 
even on brane-wave deformed backgrounds found in \cite{HY}.}
These are examples of non-integrable background geometries.
In the preceding works on chaotic strings,
only cohomogeneity-1 strings have been considered, i.e.,~some symmetries were imposed on string profiles.
The string equations of motion are then reduced to geodesic equations for particles in effective geometries~\cite{Koike:2008fs}.
Since the reduced equations of motion are given in ordinary differential equations,
 one can show the presence of chaos using standard techniques
in the field of non-linear dynamics such as Poincar\'e sections and Lyapunov spectra.

Here, we raise a simple question: 
{\it What happens to the string dynamics if such symmetries are not imposed?}
Without symmetry protection, it is necessary to take into account nonlinear fluctuations 
along the string, and further dynamics is introduced to the study of the chaotic strings.
In this paper, we address the above question focusing on the 5D AdS soliton spacetime.
We are interested in non-integrable geometries because the strings are considered 
to be non-integrable in most of the asymptotically AdS spacetimes.
We solve the full string equations of motion, given as partial differential equations.
In analyzing particle's chaos, dynamical variables depend only on time, 
but here the spatial dependence of the strings adds complexity to analysis.
The strings will show chaos in the infinite-dimensional phase space arising 
from the string's spatial direction.
We will refer such infinite-dimensional chaos as 
the turbulence in this paper. 

There are also related works on the chaos and turbulence in string theory. 
The dynamics of D0-branes can be described by matrix models \cite{BFSS,BMN}.
Their chaotic motions have been shown in \cite{chaos-BFSS,Berenstein,chaos-BMN} 
by following the procedure in classical Yang-Mills systems \cite{YM,deformed-YM}.
In the case of D7-branes, one can consider a holographic QCD setup and 
their chaotic behavior is closely related to the chiral condensate on the dual gauge-theory 
side \cite{HMY}. Since the classical chaos on the string-theory side corresponds 
to the quantum nature on the gauge-theory side, this result may imply 
a quantum analogue of the butterfly effect. As well as chaotic behaviors, 
weak turbulence on classical strings and D7 branes has also been studied 
in Refs.~\cite{Ishii:2015wua,Hashimoto:2014yza,Hashimoto:2014xta,Hashimoto:2014dda}.

This paper is organized as follows.
In section \ref{sec:adssoliton}, we introduce an AdS soliton background with the regular coordinates
and derive the equations of motion for classical strings in a conformal gauge. 
In section \ref{sec:chaos_coh1}, we use a cohomogeneity-1 string ansatz and revisit classical chaos
on the AdS soliton background, which has been studied in Ref.~\cite{Basu:2011dg}.
In section \ref{sec:chaos_string}, we include nontrivial dependence on 
the spatial direction of the string world-sheet and study the string's motion and turbulent behaviour.
Section \ref{sec:summary} is devoted to summary and discussion. 
Appendices explain some details on discretized numerical calculations. 

\section{Strings in AdS soliton}
\label{sec:adssoliton}

In this section, we shall introduce an AdS soliton background \cite{Horowitz:1998ha}
and derive the classical action to describe a string propagating in this background. 
The background is explicitly written down with certain coordinates for our use. 

\subsection{AdS soliton spacetime}

Let us consider a five-dimensional AdS soliton solution \cite{Horowitz:1998ha}. 
This can be regarded as a one-parameter deformation of the original AdS$_5$ geometry.
The metric part is given by\footnote{Here the Ramond-Ramond (R-R) five-form field strength is not
 written down because we will concentrate on the classical dynamics of the bosonic part 
of type IIB superstring theory and the R-R sector is not relevant to our analysis.} 
\begin{eqnarray}
 ds^2 &=& \frac{R^2}{z^2}\left[-dt^2+\frac{dz^2}{f(z)}+f(z)dx^2+d\vec{y}^{\,2}\right]\ , 
 \quad \vec{y}=(y_1,y_2)\ , \label{2.1} \\
  f(z) &\equiv& 1-\left(\frac{z}{z_0}\right)^4\ , \nonumber 
\end{eqnarray}
where $R$ is the AdS radius, the AdS boundary is at $z=0$, and the tip of the AdS soliton is located at $z=z_0$.
The $x$ coordinate is compactified on a circle $S^1$ with a periodicity 
$x \simeq x+ \pi z_0$ in order to avoid a conical singularity at the tip.
The metric (\ref{2.1}), however, still has a coordinate singularity at $z=z_0$ 
and may not be suitable for numerical calculations of the string dynamics.

To use a regular metric, 
we introduce the following polar coordinates in the $(z,x)$-plane: 
\begin{equation}
 r=\exp\left[-\textrm{arctanh}\left(\frac{z}{z_0}\right)
 -\textrm{arctan}\left(\frac{z}{z_0}\right)\right] \ ,\quad
 \theta=\frac{2}{z_0}x\ .
  \label{rthdef}
\end{equation}
The range of $r$ is $0\leq r < 1$, where $r=0$ and $r=1$ correspond to
the tip ($z=z_0$) and the boundary ($z=0$), respectively. 
Note here that $z$ can be expressed as a function of $r$ 
by inverting the coordinate transformation (\ref{rthdef}). (We do this numerically in our calculations.)
In terms of the new coordinates, the metric (\ref{2.1}) can be rewritten as
\begin{equation}
ds^2 = \frac{R^2}{z^2}\left[-dt^2+\frac{z_0^2f(z)}{4r^2}(dr^2+r^2d\theta^2)+d\vec{y}^{\,2}\right]\ .
\label{metric_r_theta}
\end{equation}
In this expression, the $(r,\theta)$ part of the metric is conformally flat and 
its conformal factor, which is proportional to $f(z)/r^2$,  
is regular at the tip because 
$r\sim (1-z/z_0)^{1/2}$ at $z\sim z_0$. 
The domain of $\theta$ is $0\leq \theta <2\pi$ and the conical singularity is avoided.
From the metric (\ref{metric_r_theta}), it is clear that the topology of the AdS soliton 
is given by $R_t\times R^{2}\times D_2$ where $D_2$ is a two-dimensional 
disc parametrized by $(r,\theta)$.

Since the metric~(\ref{metric_r_theta}) still has a coordinate singularity 
at the origin of the polar coordinates $r=0$, we introduce the following 
``Cartesian'' coordinates on the two-dimensional disc: 
\begin{equation}
 \chi_1=r\cos\theta\ ,\quad \chi_2=r\sin\theta\ .
\end{equation}
In terms of $\bm{\chi}=(\chi_1,\chi_2)$, the metric can be expressed as\footnote{
We use bold and arrow notations for vectors in $(\chi_1,\chi_2)$- and $(y_1,y_2)$-planes, respectively.
}
\begin{equation}
 R^{-2}ds^2 =
  F(\bm{\chi}^2)(-dt^2+d\vec{y}^{\,2}) + G(\bm{\chi}^2)d\bm{\chi}^2\ ,
  \label{AdSsolitonsimple}
\end{equation}
where we have introduced two scalar functions defined as 
\begin{equation}
F(\bm{\chi}^2)=\frac{1}{z^2}\ ,\quad G(\bm{\chi}^2)=\frac{z_0^2f(z)}{4z^2|\bm{\chi}|^2}\ . 
\label{FGfuncdef}
\end{equation}
Note here that $z$ should be regarded as a function of $r=|\bm{\chi}|$ 
by inverting the coordinate transformation (\ref{rthdef}).

In numerical calculations, we will work in units in which $z_0=1$. 
This dimensionful parameter $z_0$ can be easily retrieved in results 
whenever we want.

\subsection{The classical string action}

Let us introduce the classical action of a string propagating in the AdS soliton background. 
The classical dynamics of the fundamental string is described 
with the Nambu-Goto action, 
\begin{equation}
S=-\frac{1}{2\pi\alpha'}\int \!d^2\sigma\, \sqrt{-h}\ , \quad 
h \equiv \textrm{det}(h_{ab})\ , 
\label{NGaction}
\end{equation}
where $h_{ab}$ is the induced metric on the string. 
The prefactor corresponds to the string tension $T = 1/(2\pi\alpha')$.

Let $u$ and $v$ denote the world-sheet coordinates. 
Then the string in the target space is parametrized as
\begin{equation}
 t=t(u,v)\ ,\quad \bm{\chi}=\bm{\chi}(u,v)\ ,\quad \vec{y}=\vec{y}(u,v)\ .
\end{equation}
Plugging them in the metric (\ref{AdSsolitonsimple}), 
the components of the induced metric $h_{ab}$ are evaluated as
\begin{align}
 R^{-2} h_{uu} &=
 F(\bm{\chi}^2)(-t_{,u}^2 +\vec{y}_{,u}^{\,2})
 + G(\bm{\chi}^2)\bm{\chi}_{,u}^2\ , \\
  R^{-2} h_{vv} &=
 F(\bm{\chi}^2)(-t_{,v}^2 +\vec{y}_{,v}^{\,2})
 + G(\bm{\chi}^2)\bm{\chi}_{,v}^2\ , \\
   R^{-2} h_{uv} &=
 F(\bm{\chi}^2)(-t_{,u}t_{,v} +\vec{y}_{,u}\vec{y}_{,v})
 + G(\bm{\chi}^2)\bm{\chi}_{,u}\cdot\bm{\chi}_{,v}\ . 
 \label{gammas}
\end{align}
Using the reparametrization freedom of the world-sheet
coordinates, we impose the double null condition on the induced metric 
as follows: 
\begin{equation}
C_1\equiv h_{uu}=0\ ,\qquad
C_2\equiv h_{vv}=0\ .
\label{CON}
\end{equation}
Under the double null condition, the Nambu-Goto action (\ref{NGaction}) 
can be rewritten as
\begin{equation}
\begin{split}
S&
=-\frac{1}{2\pi\alpha'}\int\! dudv\, \sqrt{h_{uv}^2-h_{uu}h_{vv}}
 \\
 &= \frac{\sqrt{\lambda}}{2\pi}\int\! dudv\,\bigg[
 F(\bm{\chi}^2)(-t_{,u}t_{,v} +\vec{y}_{,u}\vec{y}_{,v})
 + G(\bm{\chi}^2)\bm{\chi}_{,u}\cdot\bm{\chi}_{,v}
 \bigg]\ ,
\end{split}
\label{action1}
\end{equation}
where in the second equality the double null condition
(\ref{CON}) and $h_{uv}<0$ have been utilized.
Also, the 't Hooft coupling is defined by $\lambda \equiv R^4/\alpha'{}^2$.

By taking the variations of the classical action, one can derive 
the string equations of motion: 
\begin{align}
&t_{,uv} = -\frac{F'}{F}[(\bm{\chi}\cdot \bm{\chi}_{,v})t_{,u}+(\bm{\chi}\cdot \bm{\chi}_{,u})t_{,v}] \ , \label{evol0_t}\\
&\vec{y}_{,uv} = -\frac{F'}{F}[(\bm{\chi}\cdot \bm{\chi}_{,v})\vec{y}_{,u}+(\bm{\chi}\cdot \bm{\chi}_{,u})\vec{y}_{,v}] \ , \label{evol0_y}\\
&\bm{\chi}_{,uv}=-\frac{G'}{G}[
(\bm{\chi}\cdot\bm{\chi}_{,v})\bm{\chi}_{,u}+(\bm{\chi}\cdot\bm{\chi}_{,u})\bm{\chi}_{,v}-(\bm{\chi}_{,u}\cdot\bm{\chi}_{,v})\bm{\chi}
 ]\nonumber\\
 &\hspace{6cm}
 +\frac{F'}{G}(-t_{,u}t_{,v} +\vec{y}_{,u}\vec{y}_{,v})\bm{\chi}
 \ ,
\label{evol0_x}
\end{align}
where $F'\equiv dF/d(\bm{\chi}^2)$ and $G'\equiv dG/d(\bm{\chi}^2)$.
It is easy to check that the constraint equations~(\ref{CON}) are conserved in time evolution:
$\partial_v C_1=\partial_u C_2=0$.

The evolution equations in the form of (\ref{evol0_t}-\ref{evol0_x}) are actually numerically 
unstable under time evolution. To realize stable evolution, we solve the constraints 
$h_{uu}=h_{vv}=0$ for $t_{,u}$ and $t_{,v}$ and choose the positive signature 
for the quadratic equations:
\begin{align}
t_{,u} &= \sqrt{\vec{y}_{,u}^{\,2} + H(\bm{\chi}^2)\bm{\chi}_{,u}^2} \ , \\
t_{,v} &= \sqrt{\vec{y}_{,v}^{\,2} + H(\bm{\chi}^2)\bm{\chi}_{,v}^2} \ ,
\end{align}
where $H \equiv G/F$. Taking the positive signature specifies that $\partial_u$ 
and $\partial_v$ are future directed null vectors. 
We use these conditions in (\ref{evol0_t}-\ref{evol0_x}).

In the following, it is often convenient to use orthogonal coordinates $(\tau,\sigma)$ 
defined by 
\begin{equation}
 \tau =u+v\ ,\qquad \sigma=u-v\ .
  \label{tau_sigma}
\end{equation}
In these coordinates, the string action (\ref{action1}) becomes\footnote{
When the double null condition is imposed, the Nambu-Goto action (\ref{action1}) 
takes the form of a gauge-fixed Polyakov action. The Polyakov action is given by
\begin{equation}
S=\frac{1}{4\pi \alpha'}\int\! d^2\sigma\, \sqrt{-\gamma}\, \gamma^{ab}\,
\partial_a X^\mu \partial_b X^\nu g_{\mu\nu}\ ,
\end{equation}
where $\gamma$ and $g$ are the worldsheet and spacetime metrics. 
The expressions (\ref{action1}) and (\ref{action2}) are reproduced 
when we utilize the worldsheet metric
$\gamma_{ab}d\sigma^ad\sigma^b = -2 dudv = (-d\tau^2+d\sigma^2)/2$.
}
\begin{equation}
 S=-\frac{\sqrt{\lambda}}{4\pi}\int d\tau d\sigma\,\eta^{ab}\bigg[
 F(\bm{\chi}^2)(-t_{,a}t_{,b} +\vec{y}_{,a}\vec{y}_{,b})
 + G(\bm{\chi}^2)\bm{\chi}_{,a}\cdot\bm{\chi}_{,b}
 \bigg]\ ,
 \label{action2}
\end{equation}
where $\eta_{ab}=\textrm{diag}(-1,1)$ and $a,b=\tau,\sigma$. 
The evolution and constraint equations can also be rewritten in terms of $(\tau,\sigma)$
if we replace the derivatives as $\partial_u=\partial_\tau+\partial_\sigma$ 
and $\partial_v=\partial_\tau-\partial_\sigma$.
Hereafter, 
we use $\partial_\tau\equiv {}^\cdot$ and $\partial_\sigma\equiv {}'$ for $\tau$ and $\sigma$ derivatives.

The above action is invariant under $t\to t+c_1$ and $\chi_1+i \chi_2 \to e^{ic_2}(\chi_1+i \chi_2)$
where $c_{1,2}$ are real constants. 
The Noether charges associated with these symmetries are given by 
\begin{equation}
 E=\int^{2\pi}_0 \frac{d\sigma}{2\pi}\, p_t\ ,\quad
 J=\int^{2\pi}_0 \frac{d\sigma}{2\pi}\, \bm{\chi}\times \bm{p}_\chi\ ,
 \label{Ek}
\end{equation}
where $p_t=F\,\dot{t}$
and $\bm{p}_\chi=G\,\dot{\bm{\chi}}$ are the conjugate momenta 
of $t$ and $\bm{\chi}$, respectively. 
These correspond to the energy and $(\chi_1,\chi_2)$-plane's angular momentum, 
respectively.\footnote{
From the translation symmetry in the $(y_1,y_2)$-plane, we can also obtain other conserved quantities: 
$\vec{P}=\int^{2\pi}_0 d\sigma/(2\pi)\, \vec{p}_y$ where $\vec{p}_y=F\dot{\vec{y}}$.}
Note that in the above expressions of the Noether charges the prefactor $\sqrt{\lambda}$ 
has been dropped for notational simplicity.

\section{Chaos in cohomogeneity-1 strings}
\label{sec:chaos_coh1}

The preceding work \cite{Basu:2011dg} studied the classical dynamics of strings 
in an AdS soliton background using an ansatz under which
the string equations of motion reduced to a set of 
ordinary differential equations. Then it was shown that 
the reduced system exhibited chaos. In this section, we revisit this classical chaos 
 using a regular coordinate system introduced in the previous section.

\subsection{Cohomogeneity-1 string in AdS soliton}

\begin{figure}[tbp]
\begin{center}
\includegraphics[scale=0.5]{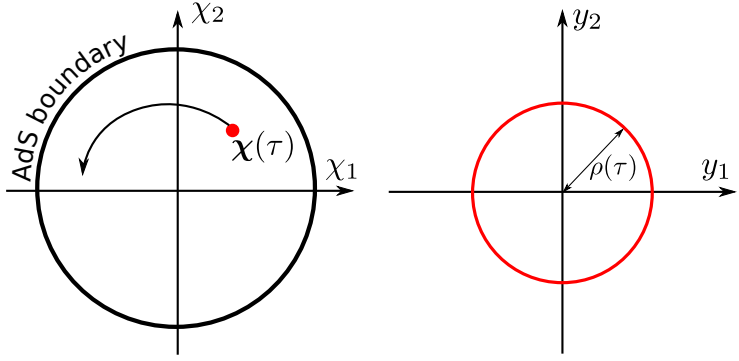}
\end{center}
 \caption{A profile of the cohomogeneity-1 string in the $(\chi_1,\chi_2)$- and $(y_1,y_2)$-planes. 
The string position/configuration is shown in red.}
\label{ponchi_col1}
\end{figure}

Let us consider the following ansatz \cite{Basu:2011dg}: 
\begin{equation}
 t=t(\tau)\ ,\quad
 \bm{\chi}=\bm{\chi}(\tau)\ ,\quad
  y_1=\rho(\tau)\cos \sigma\ ,\quad
  y_2=\rho(\tau)\sin \sigma\ .
 \label{coh1_anzatz}
\end{equation}
We will refer to string solutions with this ansatz
as cohomogeneity-1 strings, following the terminology utilized in \cite{Koike:2008fs}.
The string profile is schematically shown in Fig.~\ref{ponchi_col1}.

With this ansatz, the string action (\ref{action2}) can be rewritten as  
\begin{equation}
 S=\sqrt{\lambda}\int d\tau \left[
\frac{1}{2}F(\bm{\chi}^2)(-\dot{t}^2+\dot{\rho}^2- \rho^2)+ \frac{1}{2}G(\bm{\chi}^2)\dot{\bm{\chi}}^2\right]\ .
\label{1dLag}
\end{equation}
In the following, we drop the overall factor of $\sqrt{\lambda}$ for notational simplicity.
Our study is at the leading order in the large $\sqrt{\lambda}$ limit, and  
this overall factor can be easily recovered. 

The energy and the angular momentum are given by 
\begin{equation}
E=F(\bm{\chi}^2)\,\dot{t}\ ,\quad
 J=\bm{\chi}\times \bm{p}_\chi\ .
 \label{EJ_coh1}
\end{equation}
Then, after appropriate Legendre transformation of the Lagrangian~(\ref{1dLag}), 
we can eliminate $\dot{t}$ using the energy conservation and obtain a modified Lagrangian as
\begin{equation}
 L=\frac{1}{2}F(\bm{\chi}^2)(\dot{\rho}^2-\rho^2)+\frac{1}{2}G(\bm{\chi}^2)
\dot{\bm{\chi}}^2+\frac{E^2}{2F}\ .
\end{equation}
If we use the polar coordinates $(r,\theta)$, instead of $(\chi_1,\chi_2)$, 
$\theta(\tau)$ becomes cyclic. Hence $\theta(\tau)$ can also be eliminated 
due to the conservation of $J$.
However, the resultant equations become singular at $r=0$ and would not be suitable 
for numerical calculations if the string reaches $r=0$.
Therefore, instead of eliminating $\theta(\tau)$, we regard 
$\chi_1(\tau)$ and $\chi_2(\tau)$ as independent variables.
We introduce the conjugate momenta of $\rho$ and $\bm{\chi}$,
\begin{equation}
 p_\rho=F\dot{\rho}\ ,\quad \bm{p}_\chi=G \dot{\bm{\chi}}\ .
\label{conjm}
\end{equation}
The Hamiltonian is obtained as
\begin{equation}
 H=\frac{p_\rho^2-E^2}{2F(\bm{\chi}^2)}+\frac{\bm{p}_\chi^2}{2G(\bm{\chi}^2)}+\frac{1}{2}F(\bm{\chi}^2)\rho^2\ .
\label{Ham}
\end{equation}
The Hamilton equations are given by 
\begin{equation}
\dot{p}_\rho=-F \rho\ ,\qquad
 \dot{\bm{p}}_\chi=
 \left[
\frac{F'}{F^2}(p_\rho^2-E^2)+\frac{G'}{G^2}\bm{p}_\chi^2-F'\rho^2
 \right]\bm{\chi} \ ,
 \label{eom_coh1}
\end{equation}
together with the conjugate momenta (\ref{conjm}). 
The constraints (\ref{CON}) lead to the Hamiltonian constraint $H=0$,
and this should be imposed on initial conditions.

\subsection{Poincar\'e section and Lyapunov exponent}
\label{PoinLyap}

For initial conditions to solve the Hamilton equations, 
we set $\chi_2=p_\rho=p_{\chi_1}=0$ for simplicity and regard $(E,J,\chi_1)$ as free parameters.
Using the constraint $H=0$ and the second equation in (\ref{EJ_coh1}),
the initial values of $\rho$ and $p_{\chi_2}$ can be expressed in terms of $(E,J,\chi_1)$ as
\begin{equation}
 p_{\chi_2}=\frac{J}{\chi_1}\ ,\quad
  \rho^2=\frac{1}{F(\chi_1^2)^2}\left(E^2-\frac{F(\chi_1^2)}{\chi_1^2G(\chi_1^2)}J^2\right)\ .
\end{equation}
Note that the positivity of the second equation also gives a constraint among possible combinations of $(E,J,\chi_1)$.
After solving the equations of motion in terms of $\bm{\chi}, \, \rho, \, \bm{p}_\chi$ and $p_\rho$, 
we can compute the radial coordinate $r=|\bm{\chi}|$ and its conjugate momentum $p_r\equiv G \dot{r}$.
As explained in the previous subsection, the angular variable $\theta(\tau)$ is reducible.
Therefore, the phase space of the current system is 4 dimensions spanned by $(\rho,p_\rho,r,p_r)$.
Once we fix the initial conditions, the string motion in the phase space is 
constrained in a three-dimensional subspace satisfying the constraint $H=0$.

\begin{figure}
  \centering
  \subfigure[$E=1.05$, $J=0.5$]
  {\includegraphics[scale=0.45, angle=270]{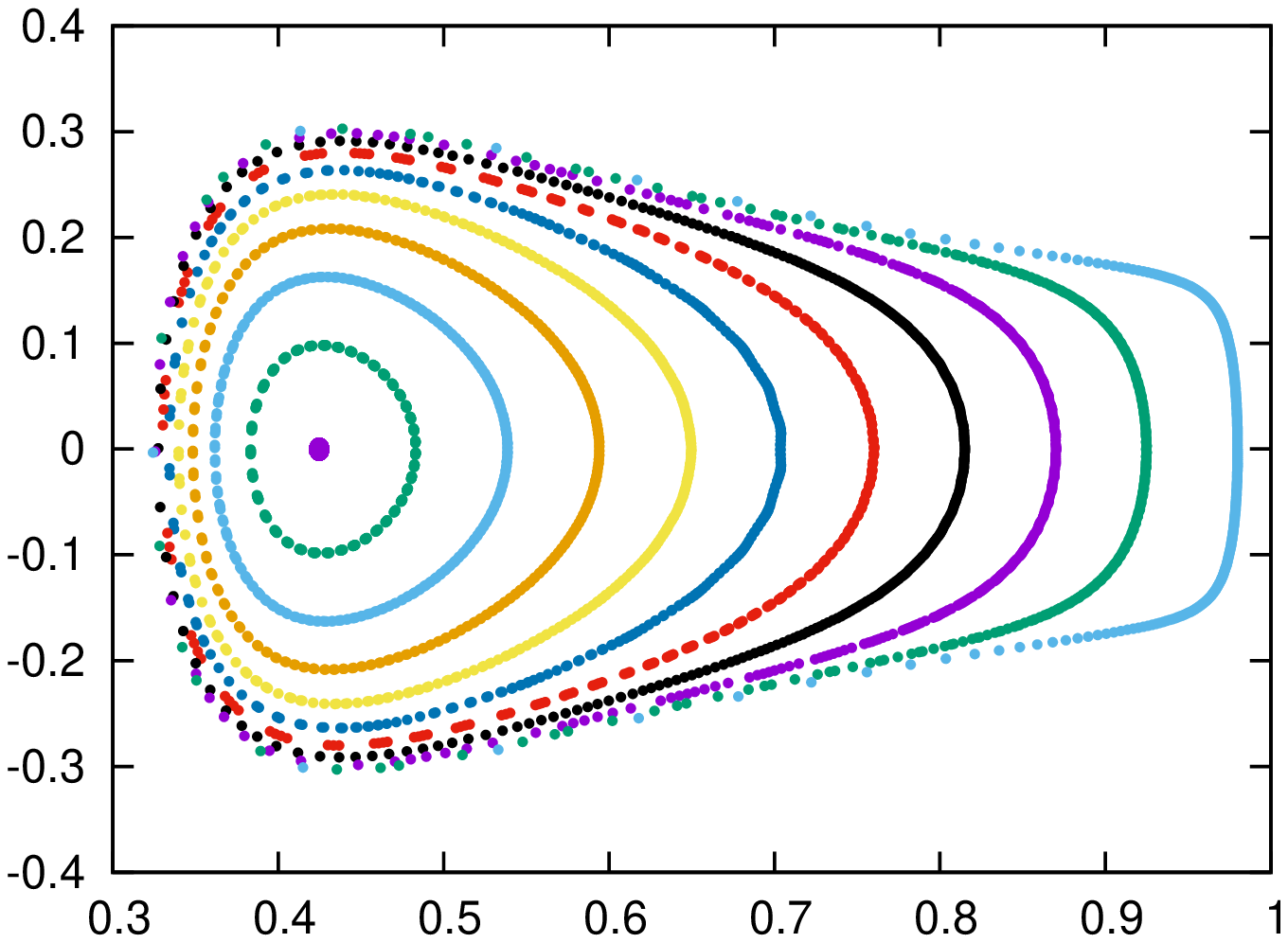}\label{poin1}
 }
 \subfigure[$E=1.1$, $J=0.5$]
  {\includegraphics[scale=0.45, angle=270]{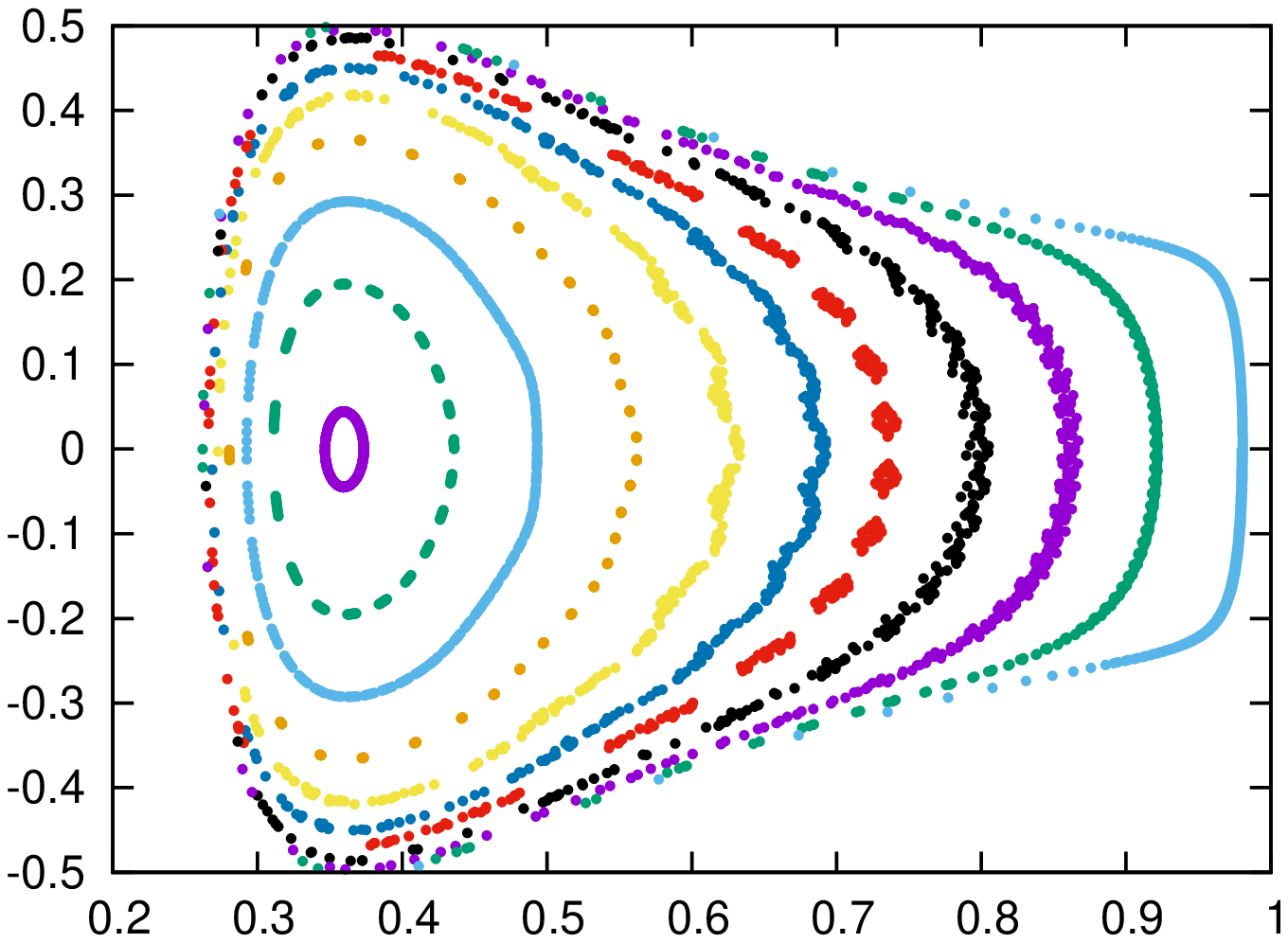}\label{poin2}
 }
 \subfigure[$E=1.2$, $J=0.5$]
  {\includegraphics[scale=0.45, angle=270]{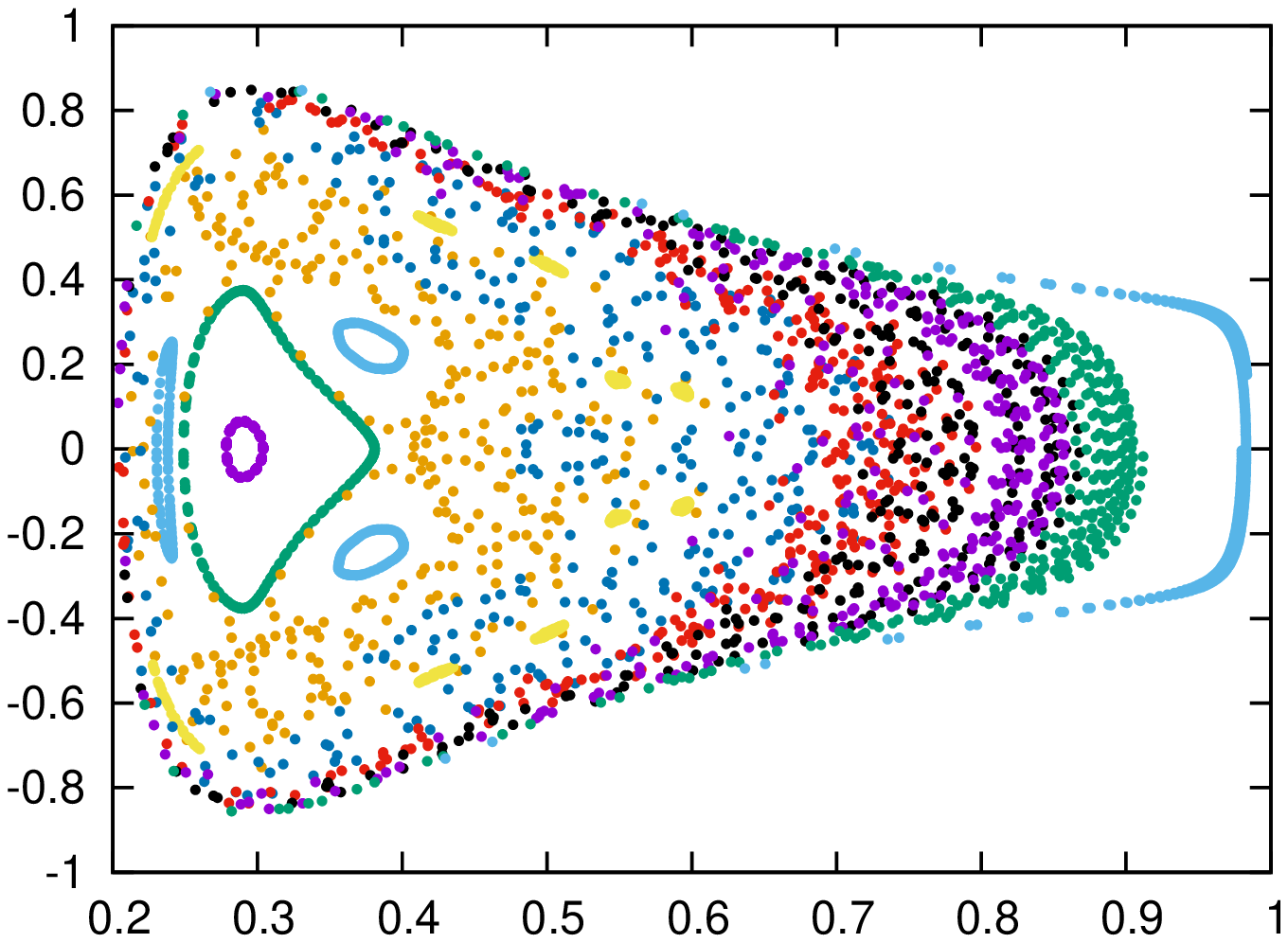}\label{poin3}
 }
  \subfigure[$E=4$, $J=0.5$]
  {\includegraphics[scale=0.45, angle=270]{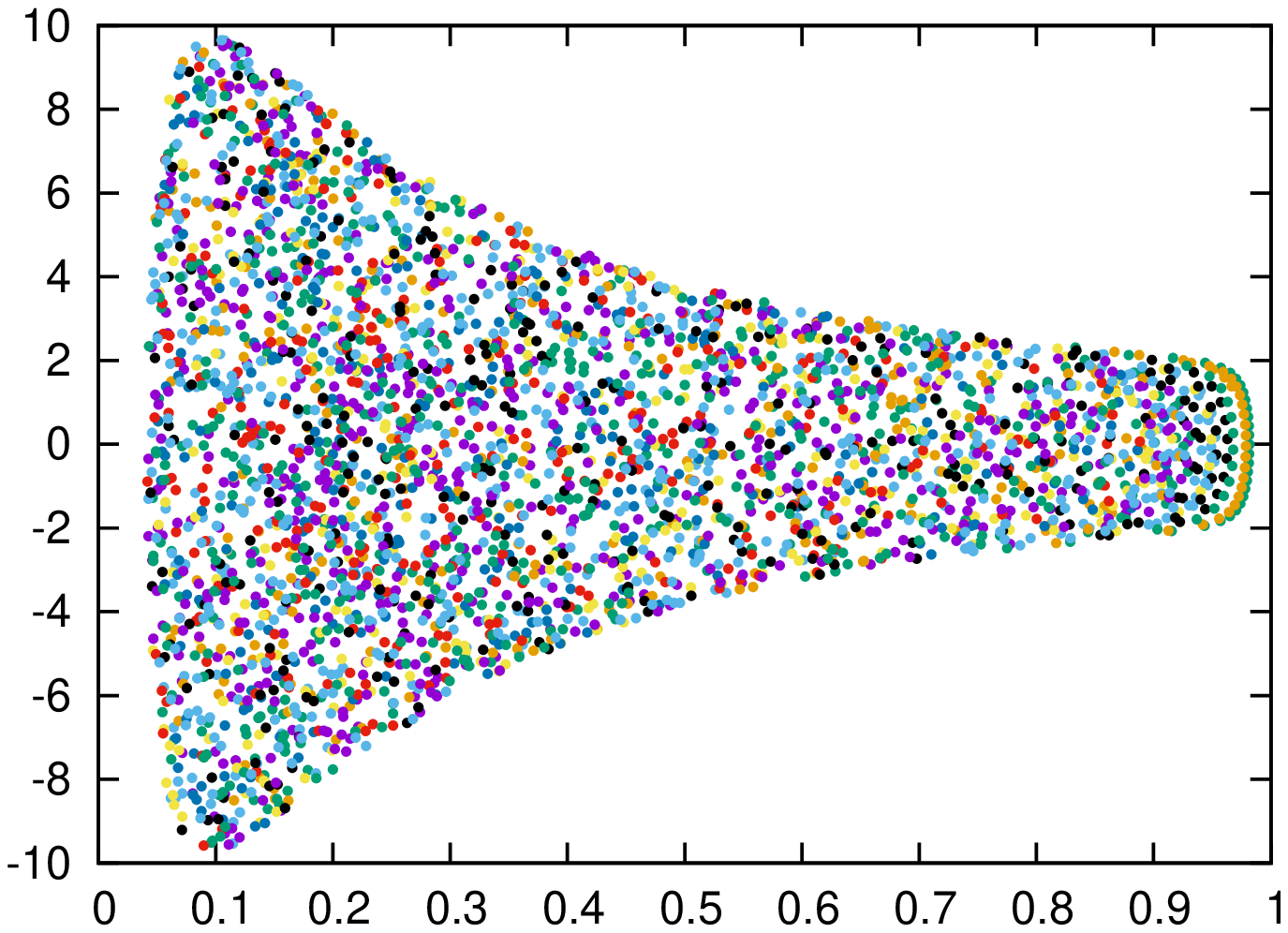}\label{poin4}
  }
  \caption{%
 Poincar\'e sections $(\rho=0)$.
 The horizontal and vertical axes are $r$ and $p_r$, respectively.
 Different colors correspond to different initial values of $\chi_1$.
 As the energy increases, the chaotic region dominates the phase space.
 }
 \label{poins}
\end{figure}

Starting from such initial conditions, we solve Eqs.~(\ref{conjm}) and (\ref{eom_coh1}) 
by using the fourth order Runge-Kutta method.
We then look at a Poincar\'e section $\rho=0$ to demonstrate chaotic motion.
In Figs.~\ref{poin1}-\ref{poin4}, 
we show $(r,p_r)$-plane plots of intersecting points of the phase space orbit 
and the Poincar\'e section. In each figure, $E$ and $J$ are fixed and the initial $\chi_1$ is varied.
Points with different colors correspond to different initial $\chi_1$.
From Fig.~\ref{poin1} to \ref{poin4},
we increase the energy while the angular momentum is fixed to $J=0.5$.
For the $E=1.05$ case, only Kolmogorov-Arnold-Moser (KAM) tori \cite{Ko,Ar,Mo} appear 
and there is no sign of chaos. 
At $E=1.1$, however, the KAM tori are destroyed gradually. 
With $E=1.2$, most of the KAM tori are destroyed and there appears the sea of chaos containing 
the surviving KAM tori as small islands. With $E=4$, the phase space is completely dominated 
by the chaotic motion. We have checked that, in $E\gtrsim 2$, the Poincar\'e section is 
qualitatively similar to the $E=4$ case. 
These results rule out the integrability of the string motion in the 5D AdS soliton.

\begin{figure}
  \centering
  \subfigure[Lyapunov spectrum]
  {\includegraphics[scale=0.45]{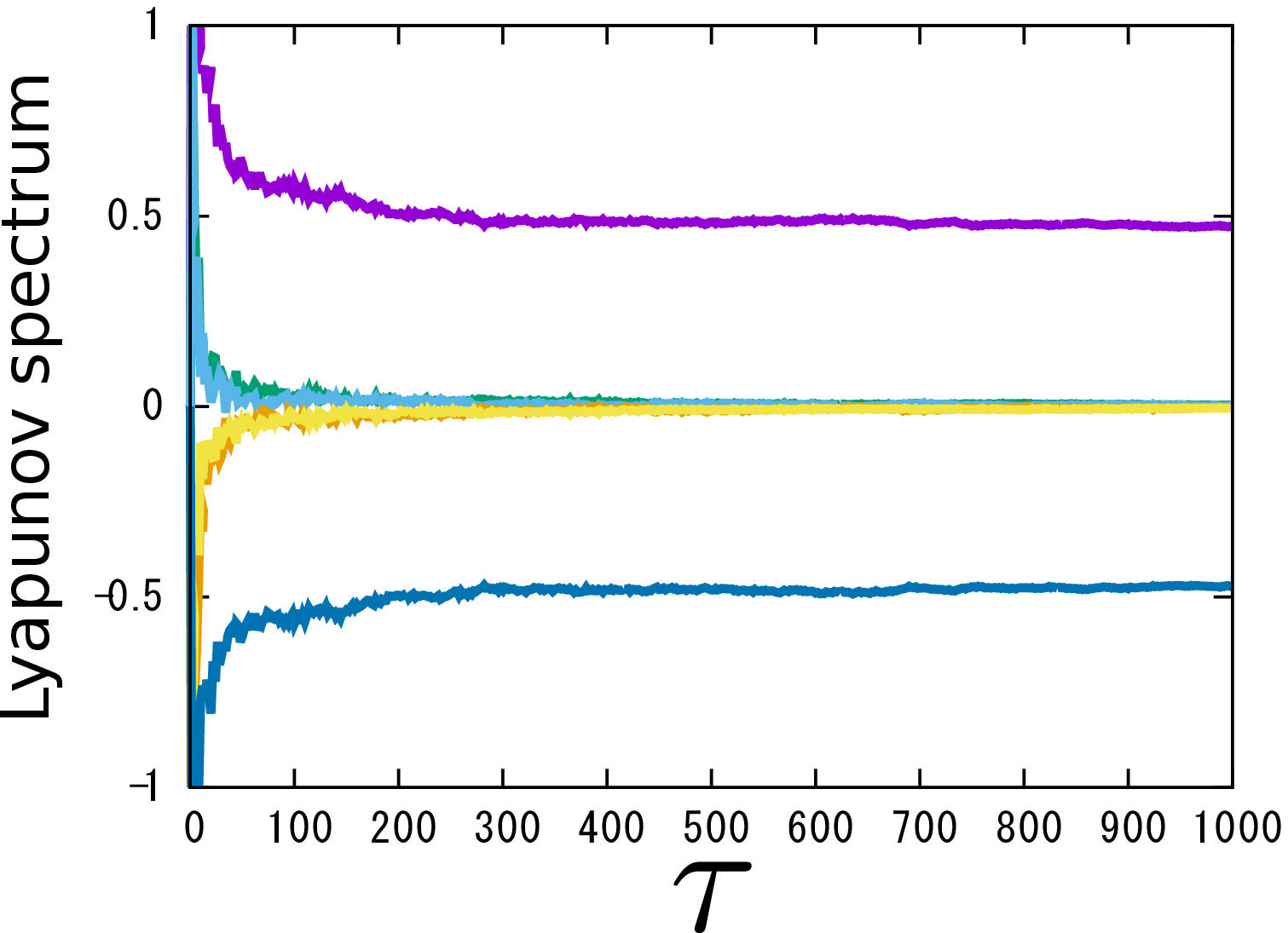}\label{Lyap_t}
 }
 \subfigure[Maximal Lyapunov exponents]
  {\includegraphics[scale=0.45]{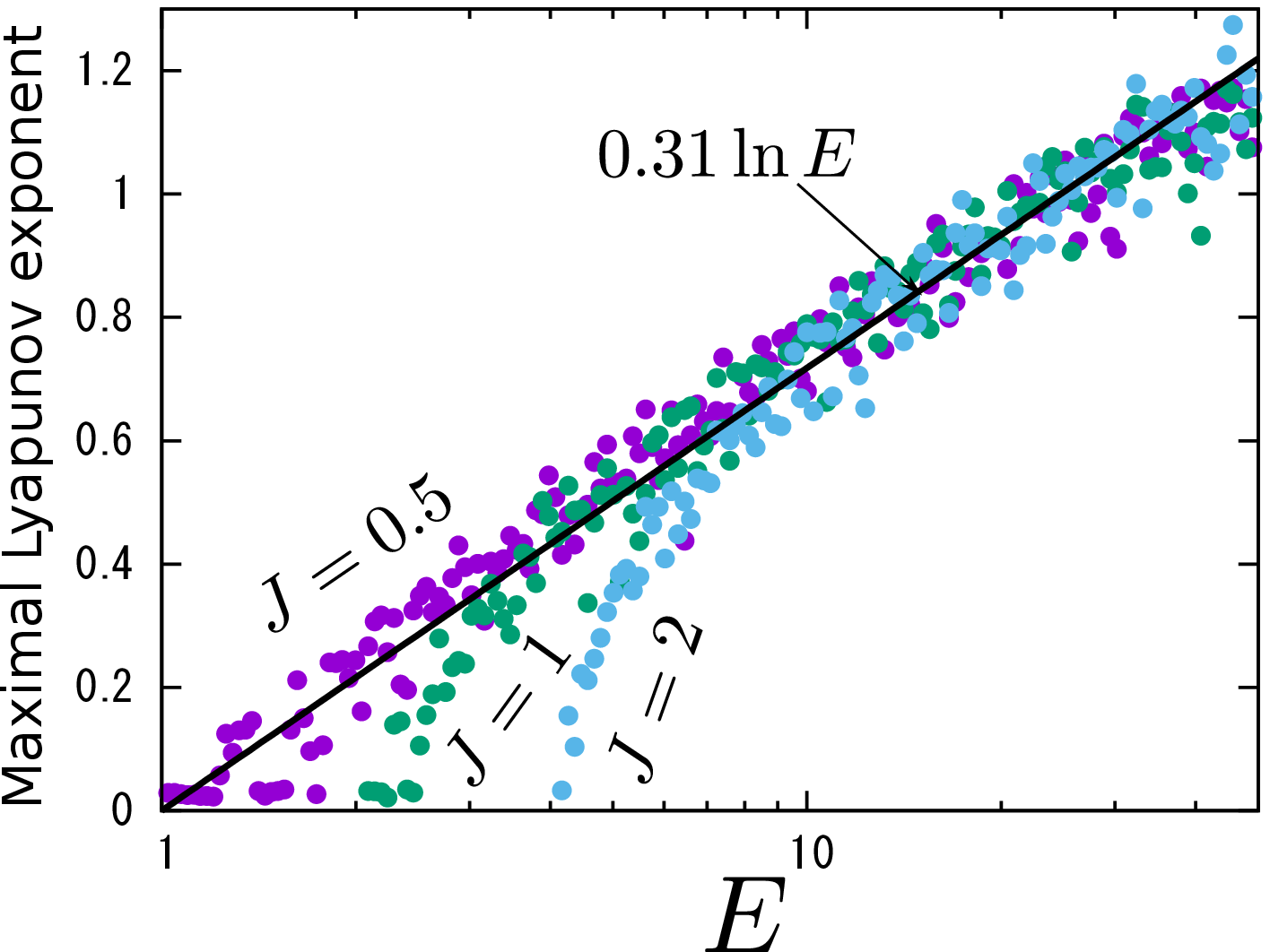}\label{Lyap_E}
 }
 \caption{%
 (a)~Convergence of the Lyapunov spectrum for $E=4$ and $J=0.5$.
 (b)~The maximal Lyapunov exponents as functions of $E$ for $J=0.5,1,2$.
 In these figures, we set $\chi_1=0.4$ and $\chi_2=p_{\chi_1}=p_\rho=0$ in the initial conditions.
 }
\end{figure}

We also compute Lyapunov exponents to evaluate the strength of the chaos.
We denote by $\bm{X}=(\bm{\chi},\rho,\bm{p}_\chi,p_\rho)$ a solution in the phase space
and consider its linear perturbation: $\bm{X}\to \bm{X}(\tau)+\bm{\delta}(\tau)$.
If $\bm{X}(\tau)$ is a chaotic solution, 
the perturbation grows exponentially as $|\bm{\delta}(\tau)|\propto e^{L \tau}$, where
the exponent $L$ is called the Lyapunov exponent.
This growth reflects the sensitivity of the time evolution to initial conditions in chaotic systems.
In $N$-dimensional phase space, there are $N$ Lyapunov exponents depending on 
perturbations for $\bm{\delta}$, and the set of the Lyapunov exponents 
$\{L_1,L_2,\cdots,L_N\}$ is called the Lyapunov spectrum.
The largest one in the spectrum is called the maximal Lyapunov exponent.
In Fig.~\ref{Lyap_t}, we show the Lyapunov spectrum as functions of 
$\tau$: $L_k\sim \ln |\bm{\delta}(\tau)|/\tau $. 
We have used the Shimada-Nagashima method~\cite{Shimada},
which is a standard numerical method to obtain the Lyapunov spectrum 
(See also the appendix to Ref.~\cite{chaos-BMN}).
We see that one of the exponents approaches a positive value $L=0.473$. 
In Fig.~\ref{Lyap_E}, we show the maximal Lyapunov exponents as functions of $E$ for $J=0.5,1,2$.
We set $\chi_1=0.4$ and $\chi_2=p_{\chi_1}=p_\rho=0$ in the initial conditions.
The maximal Lyapunov exponents approach a linear function of $\ln E$ in large $E$.
The slope does not seem to depend on $J$. Fitting the plots, we obtain $L\simeq 0.31 \ln E$.
These results explicitly demonstrate that the string dynamics in the AdS soliton background admits
sensitivity to initial conditions.

\section{Turbulent string condensation}
\label{sec:chaos_string}

In the previous section, an ansatz has been used to remove the $\sigma$-coordinate 
dependence from the string dynamics, and the string equations of motion were reduced to ODEs.
In this section, we do not use such an ansatz but solve the full evolution equations given 
as partial differential equations. Our aim is to consider how a system depending on 
more than one variables inherits the particle chaos studied in the previous section. 
We will argue it is the turbulent behaviour in the string dynamics that becomes relevant.

\subsection{String motion}

\begin{figure}
  \centering
  \subfigure[$0\leq t\leq 3.6$]
  {\includegraphics[scale=0.42]{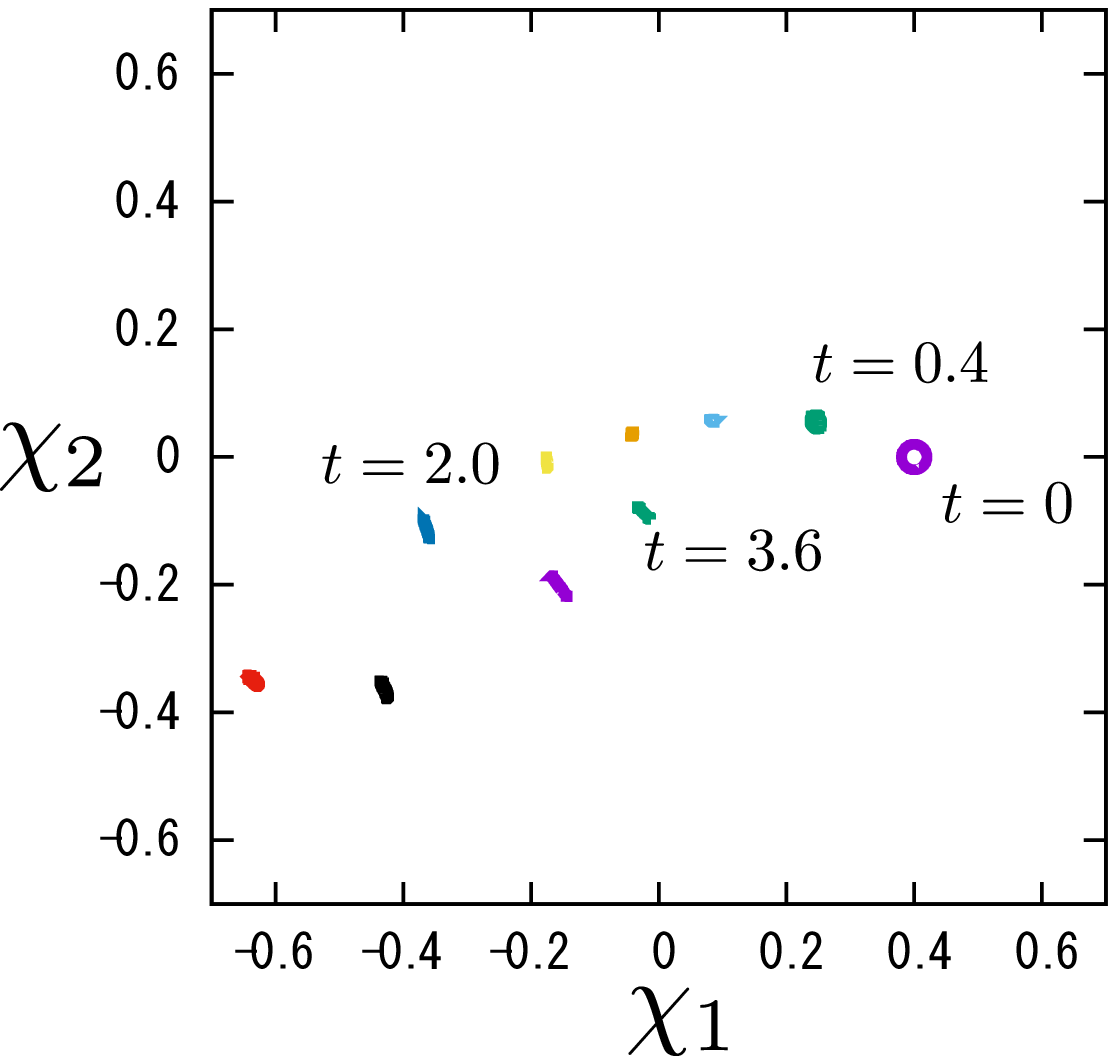}\label{profile1}
 }
 \subfigure[$7.8\leq t\leq 10.6$]
  {\includegraphics[scale=0.42]{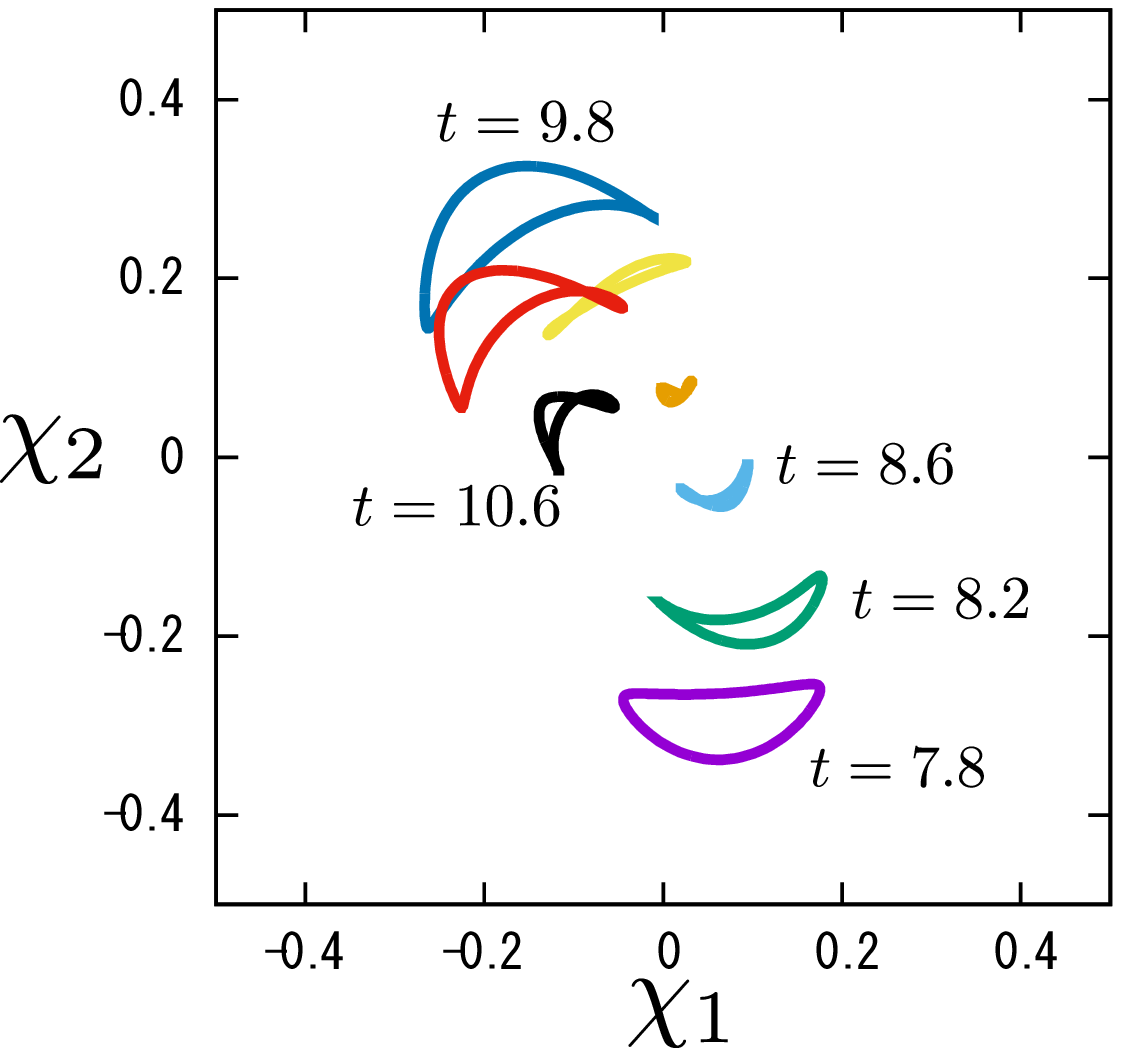}\label{profile2}
 }
 \subfigure[$50\leq t\leq 52.4$]
  {\includegraphics[scale=0.42]{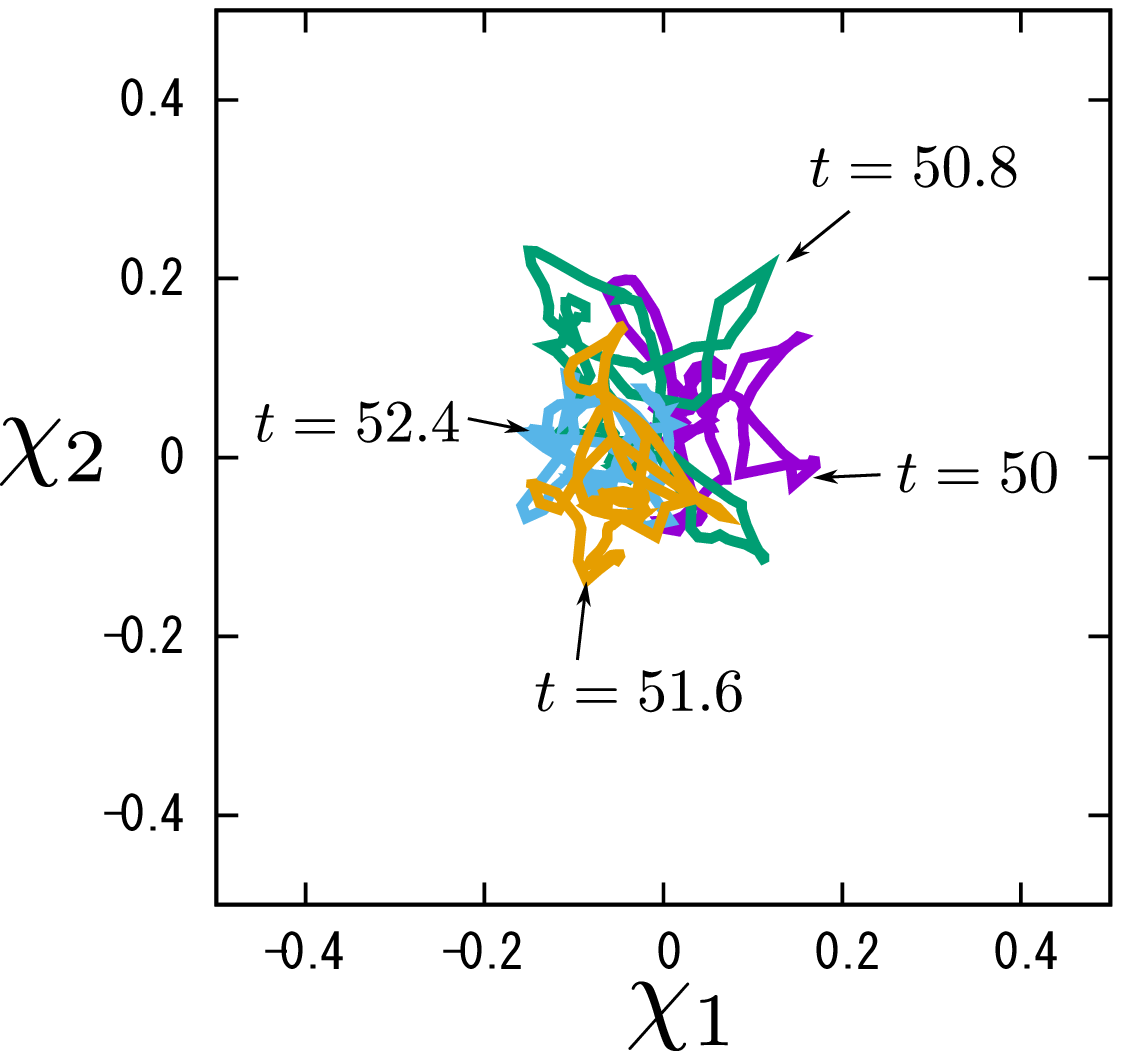}\label{profile3}
 }
  \caption{%
 Time evolution of the string in $(\chi_1,\chi_2)$-plane for $0\leq t\leq 3.6$, $7.8\leq t\leq 10.6$, and $50\leq t\leq 52.4$.
 We set the parameters in the initial data as $(\epsilon,r_0,\omega,\rho_0)=(0.02,0.4,0.4289,0.7939)$ 
 and took $z_0=1$.
 }
 \label{profile}
\end{figure}

Let us solve the string's evolution equations (\ref{evol0_t}-\ref{evol0_x}) with initial data 
that involves general dynamics of the string.
In the previous section, the string was pointlike in the $(\chi_1,\chi_2)$-plane.
Here, we consider an initial configuration in which the string is extended to a small circle as
\begin{equation}
 \chi_1|_{\tau=0}=r_0+\epsilon \cos\sigma\ ,\quad
 \chi_2|_{\tau=0}=\epsilon\sin\sigma\ .
  \label{chiInit}
\end{equation}
We set the other variables to the same initial profile as the cohomogeneity-1 string:
\begin{equation}
 t|_{\tau=0}=0\ ,\quad y_1|_{\tau=0}=\rho_0 \cos\sigma\ ,\quad y_2|_{\tau=0}
 =\rho_0 \sin\sigma\ .
  \label{tyinit}
\end{equation}
One can find an example of the initial string profile at the $t=0$ configuration in Fig.~\ref{profile1}.
The initial velocity is taken as follows: 
\begin{equation}
 \dot{\chi}_1|_{\tau=0}=-\omega \chi_2|_{\tau=0}\ ,\quad
 \dot{\chi}_2|_{\tau=0}=\omega \chi_1|_{\tau=0}\ .
\end{equation}
That is, we give an initial angular velocity to the string in the $(\chi_1,\chi_2)$-plane 
while keeping the circular configuration in Eq.~(\ref{chiInit}). Note that, while the initial data 
is given in this way, the exact circular configuration is not preserved in the time evolution.
The initial $\dot{t}$, $\dot{y}_1$ and $\dot{y}_2$ are determined by solving 
the constraint equations. See Appendix~\ref{app:init} for details of the initial data construction.
The initial data is specified by four parameters $\epsilon$, $r_0$, $\omega$ and $\rho_0$.
As a numerical scheme to solve the evolution equations, we use the method developed 
in Refs.~\cite{Ishii:2014paa,Ishii:2015wua,Ishii:2015qmj}.

We start from considering a parameter set corresponding to a chaotic situation.
In the previous section, we found that the cohomogeneity-1 string with $E=4$ and $J=0.5$ 
was chaotic (See Figs.~\ref{poin4} and \ref{Lyap_t}).  
Here, we set the parameters as $(\epsilon,r_0,\omega,\rho_0)=(0.02,0.4,0.4289,0.7939)$.
Then from (\ref{Ek}), the energy and the angular momentum are $E=4.00$ and $J=0.500$. 

Figure \ref{profile} shows snapshots of the string in the $(\chi_1,\chi_2)$-plane 
at several stages of the time evolution. In the beginning $0\leq t\leq 3.6$,
the string motion is similar to that of the cohomogeneity-1 string.
The string size is as small as the initial configuration in the $\bm{\chi}$-plane.
In intermediate times $7.8\leq t\leq 10.6$, the string size gets larger. 
This behaviour is naturally understood from the chaos found in Sec.~\ref{sec:chaos_coh1}: 
Time evolution in chaotic systems is sensitive to a tiny difference in initial conditions, 
and because of this nature the trajectories of each string segment 
(expanded initially by $\epsilon$) tend to spread in the Lyapunov timescale.
It seems interesting that the string has the horseshoe-like profile in the $\bm{\chi}$-plane.
This would be reflecting the baker's transformation in chaotic systems. 
Different from the cohomogeneity-1 strings, the expanded string is affected by its tension in the $\bm{\chi}$-plane.
In later time, $50\leq t\leq 52.4$, the expansion of the string is 
saturated by its tension, and the string profile is jumbled up.
Once the string configuration reaches this stage, it does not seem to come back to 
``smooth'' configurations appeared in early times.

We can schematically understand that this behaviour stems from the finite size string 
without symmetry protection. Once the string has a finite size in the $\bm{\chi}$-plane, 
it can have ``internal spin.'' ``Orbital angular momentum'' can decrease if it is transferred 
to ``spin angular momentum'' (We will define and discuss the spin and orbital angular 
momenta shortly in Sec.~\ref{sec:Angdis}). Losing the orbital angular momentum, 
the string tends to stay around the center of the $\bm{\chi}$-plane.
In the rest of this section, we will study in detail that the mechanism causing this transfer 
is the turbulence phenomenon on the string \cite{Ishii:2015wua}.
For this reason, we will refer this behaviour of the string as {\it turbulent string condensation}.

\subsection{Lyapunov exponents toward turbulent string condensation}
\label{sec:string_lyapunov}

In Sec.~\ref{PoinLyap}, we have evaluated the Lyapunov exponents of the cohomogeneity-1 
string, where a symmetry was imposed as Eq.~(\ref{coh1_anzatz}). Let us consider here Lyapunov 
exponents in a more general situation in which fluctuations that do not respect that symmetry 
is introduced. If any symmetry on the string dynamics is not imposed, 
the phase space concerned with us is infinite dimensional, and
the Shimada-Nagashima method~\cite{Shimada} may not be suitable. 
Instead of it, we will use a short-cut method to roughly estimate the Lyapunov exponents as follows.

Let us consider slightly different initial conditions I and II given by
\begin{enumerate}
\renewcommand{\labelenumi}{\Roman{enumi}.}
\item $(\epsilon,r_0,\omega,\rho_0)=(0,0.4,0.4289,0.7939)$
\item $(\epsilon,r_0,\omega,\rho_0)=(0.02,0.4,0.4289,0.7939)$
\end{enumerate}
With the initial condition I, the string is cohomogeneity-1 and point-like in the $\bm{\chi}$-plane.
The condition II is the same as that for Fig.~\ref{profile}, and we find that this case results 
in the turbulent string condensation. Let $\bm{\chi}^\textrm{I}(\tau)$ and 
$\bm{\chi}^\textrm{II}(\tau,\sigma)$ denote the string solutions 
in the $\bm{\chi}$-plane for the initial conditions I and II, respectively 
(Note that the cohomogeneity-1 solution $\bm{\chi}^\textrm{I}$ does not depend on $\sigma$). 
Let us consider Fourier transformation of $\bm{\chi}^\textrm{II}(\tau,\sigma)$ 
along the $\sigma$-direction as
\begin{equation}
 \bm{\chi}^\textrm{II}(\tau,\sigma) =\sum_{n=-\infty}^\infty \bm{\chi}_n^\textrm{II}(\tau) e^{in\sigma}\ ,
\end{equation}
and define $\delta_n(\tau)$ ($n=0,1,2,\cdots$) as
\begin{equation}
 \delta_0(\tau)\equiv |\bm{\chi}^\textrm{II}_0(\tau)-\bm{\chi}^\textrm{I}(\tau)|\ ,\quad
 \delta_n(\tau)\equiv |\bm{\chi}_n^\textrm{II}(\tau)|\quad (n \geq 1)\ .
\end{equation}
That is, $\delta_0(\tau)$ measures the deviation of the ``bulk motion'' from the cohomogeneity-1 
case, and $\delta_{n\geq 1}(\tau)$ the growth of the ``internal structure'' of the string.

\begin{figure}
\begin{center}
\subfigure[$E=4.0$, $J=0.5$]{\includegraphics[scale=0.5]{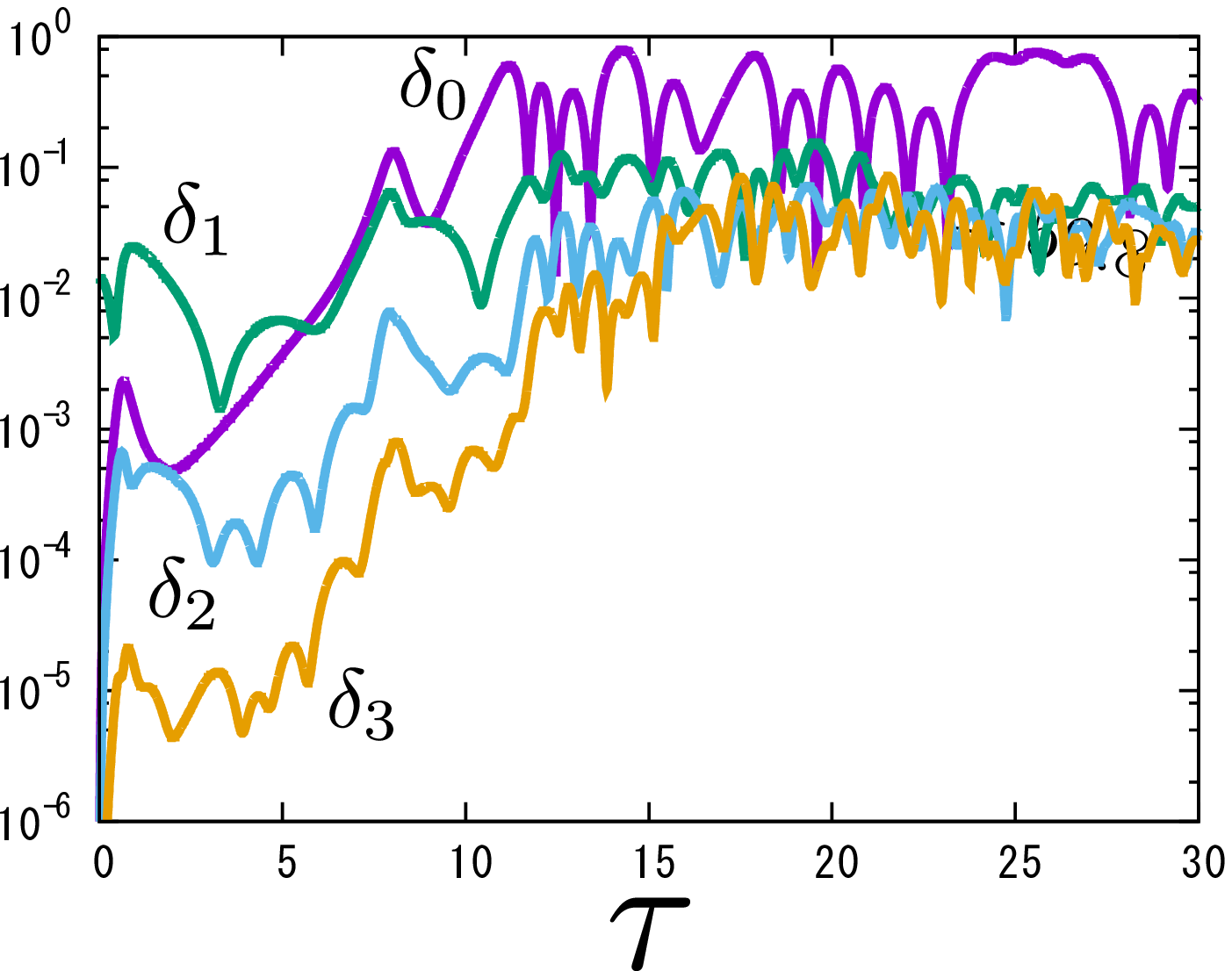}\label{deltasE4}}
\subfigure[$E=1.11$, $J=0.5$]{\includegraphics[scale=0.5]{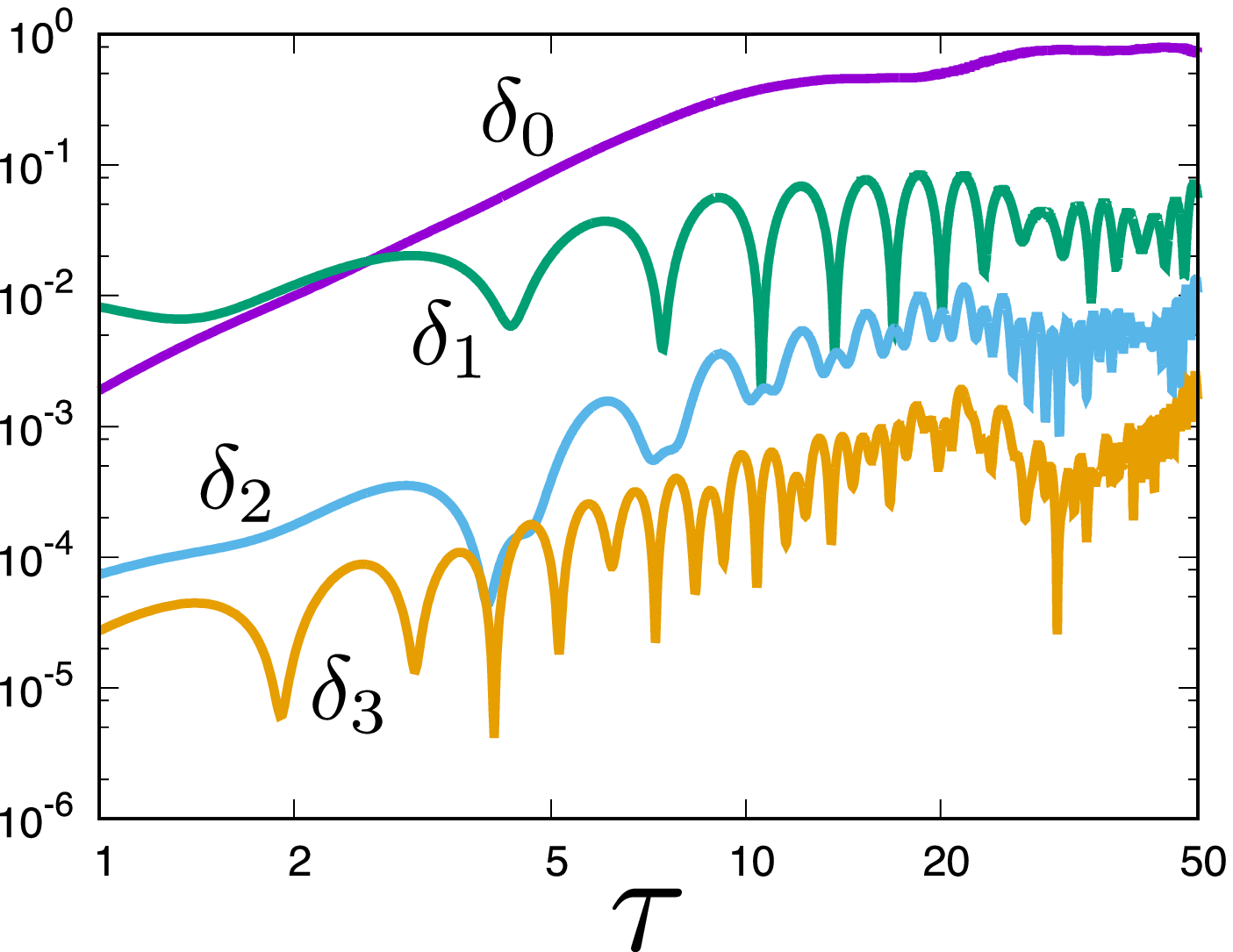}\label{delta_power}}
\end{center}
 \caption{%
Sensitivity to initial conditions of each Fourier mode.
}
\label{deltas}
\end{figure}

In Fig.~\ref{deltasE4}, we show $\delta_n$ $(n=0,1,2,3)$ as functions of $\tau$. 
In early times $\delta_n$ increase exponentially, and around at $\tau=10$ their magnitude 
get saturated due to non-linear effects. For the $n=0$ mode, the initial increase indicates 
the deviation between the trajectory of the initial condition I and that of the average string position 
of the condition II, while for $n \ge 1$, $\delta_n$ measures the excitation of modes on the string.
Fitting $\delta_n(\tau)$ with $\sim \exp(L_n t)$ in $5 \le \tau \le 10$, 
we obtain the (local) Lyapunov exponents as 
$L_0=0.66$, $L_1=0.47$, $L_2=0.55$, and $L_3=0.75$. 
These non-zero Lyapunov exponents of the symmetry breaking modes can be regarded as
the origin of the turbulent string condensation. That is, the chaotic nature known 
in the case of the cohomogeneity-1 string appears as the exponential growth of 
the symmetry breaking $n \ge 1$ modes in general string dynamics 
(that is not reduced to the ODEs).

In non-chaotic cases, in contrast, we do not find such exponential behaviours 
in $\delta_n(\tau)$. 
Choosing the input parameters as $(\epsilon,r_0,\omega,\rho_0)=(0.02,0.4,0.4289,0.06)$ 
leads to $E=1.11$ and $J=0.5$. The corresponding cohomogeneity-1 string is not 
in the chaotic region (See Fig.~\ref{poin2}). The string in the $(\chi_1,\chi_2)$-plane 
travels with keeping its initial size and does not jumble. In Fig.~\ref{delta_power}, $\delta_n$ 
are plotted in a log-log scale. It is also found that $\delta_n$ grows with power law.
This behaviour continues until $\tau \sim 20$ and then $\delta_n$ is saturated.
For $n \ge 2$ modes, the envelope of $\delta_n(\tau)$ in $\tau < 20$ behaves as $\sim \tau^{1.4}$.
For the $n=1$ mode, we find $\sim \tau^{0.7}$, but apparently this smaller exponent is due to the initial condition (\ref{chiInit}) where $\delta_1(\tau=0)$ is already of the size of $\epsilon$.

\subsection{Angular momentum distribution}
\label{sec:Angdis}

We shall study the turbulent behaviour in the string condensation quantitatively
using the angular momentum spectrum defined in the following.

Let us first decompose $\bm{\chi}$ and $\bm{p}_\chi$
into Fourier modes along the $\sigma$-direction as
\begin{equation}
 \bm{\chi}(\tau,\sigma) =\sum_{n=-\infty}^\infty \bm{\chi}_n(\tau) e^{in\sigma}\ ,\quad
 \bm{p}_\chi(\tau,\sigma) =\sum_{n=-\infty}^\infty \bm{p}_n(\tau) e^{in\sigma}\ .
\end{equation}
By substituting the above expressions into Eq.~(\ref{Ek}), 
the angular momentum spectrum $J_n$ can be obtained as
\begin{equation}
 J=\sum_{n=0}^\infty J_n\ ,\qquad
 J_0=\bm{\chi}_0 \times \bm{p}_0\ ,\quad
 J_n=2\,\textrm{Re}(\bm{\chi}_n \times \bm{p}_n^\ast)\quad (n\geq 1)\ .
\end{equation}
The spectrum is divided into the orbital angular momentum $L$ and 
the spin angular momentum $S$ as
\begin{equation}
 L=J_0\ ,\qquad S=\sum_{n=1}^\infty J_n\ .
\end{equation}
In the case of the cohomogeneity-1 string, the spin angular momentum is zero 
because $\bm{\chi}$ does not depend on $\sigma$.

\begin{figure}
  \centering
  \subfigure[$E=4.0$, $J=0.5$]
  {\includegraphics[scale=0.5]{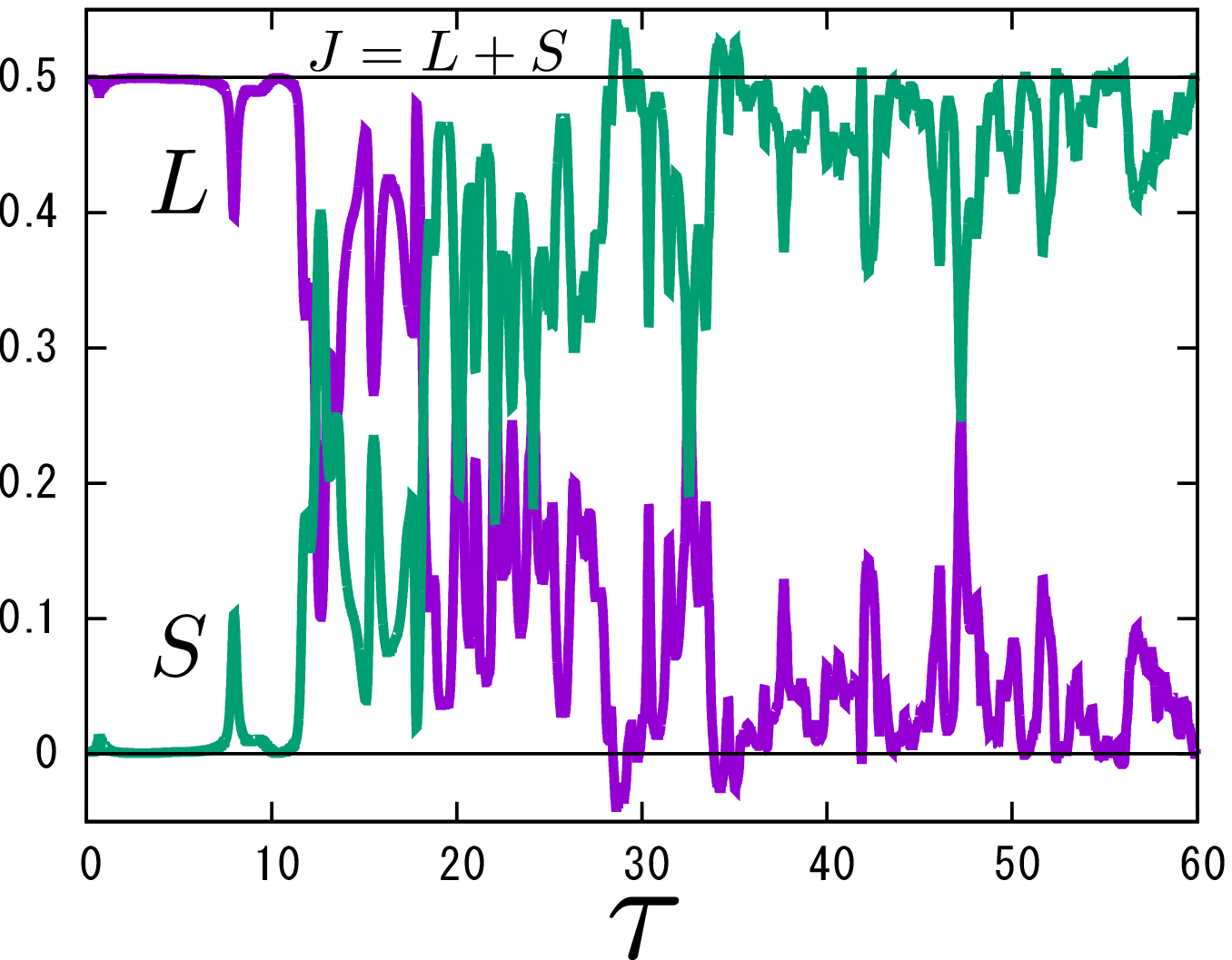}\label{LS1}
 }
 \subfigure[$E=1.11$, $J=0.5$]
  {\includegraphics[scale=0.5]{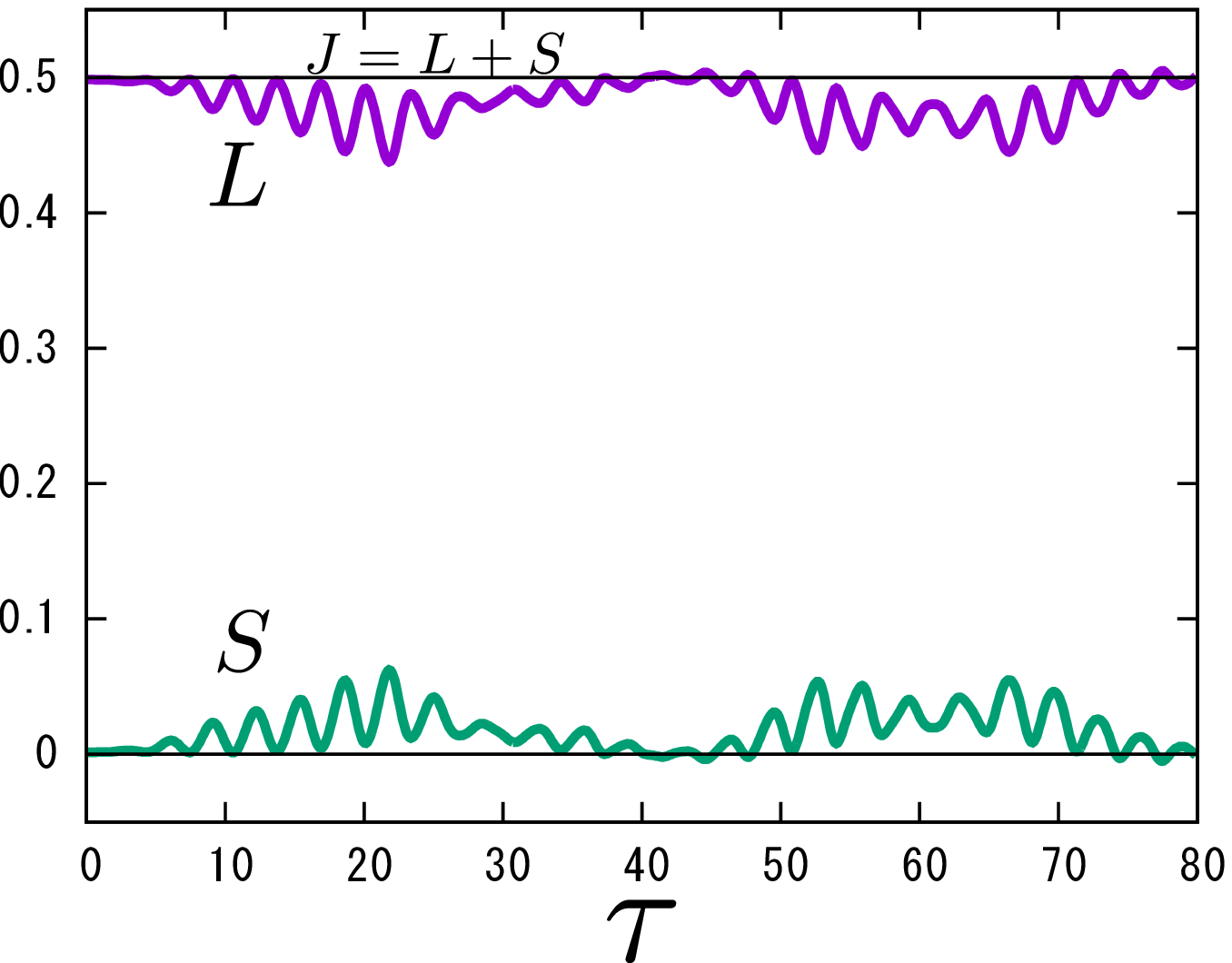}\label{LS2}
 }
  \caption{%
 Time evolution of $L$ and $S$. In the left panel, the initial data is the same as that 
 for Fig.~\ref{profile}. In the right panel, 
 we set $(\epsilon,r_0,\omega,\rho_0)=(0.02,0.4,0.4289,0.06)$.
 }
 \label{LandS}
\end{figure}

In Fig.~\ref{LandS}, we plot the $\tau$-dependence of $L$ and $S$ for two different initial data.
Figure \ref{LS1} corresponds to the string solution shown in Fig.~\ref{profile}.
Initially, the total angular momentum is dominated by the the orbital angular momentum 
inherited from the initial configuration as seen in the string motion in Fig.~\ref{profile1}. 
At late times, we see that the spin contribution becomes large and eventually
dominates the total angular momentum. It may be possible to regard the spin angular momentum as
an order parameter for the turbulent string condensation. Figure \ref{LS2} is for a smaller energy 
without chaotic behaviour: $E=1.11$ and $J=0.50$. The parameter choice is 
$(\epsilon,r_0,\omega,\rho_0)=(0.02,0.4,0.4289,0.06)$ and this has also been considered 
at the end of the last subsection. We find that the angular momentum is always dominated 
by the orbital one. There is no turbulent string condensation in this case.

\begin{figure}
  \centering
  \subfigure[$E=4.0$, $J=0.5$]
  {\includegraphics[scale=0.45]{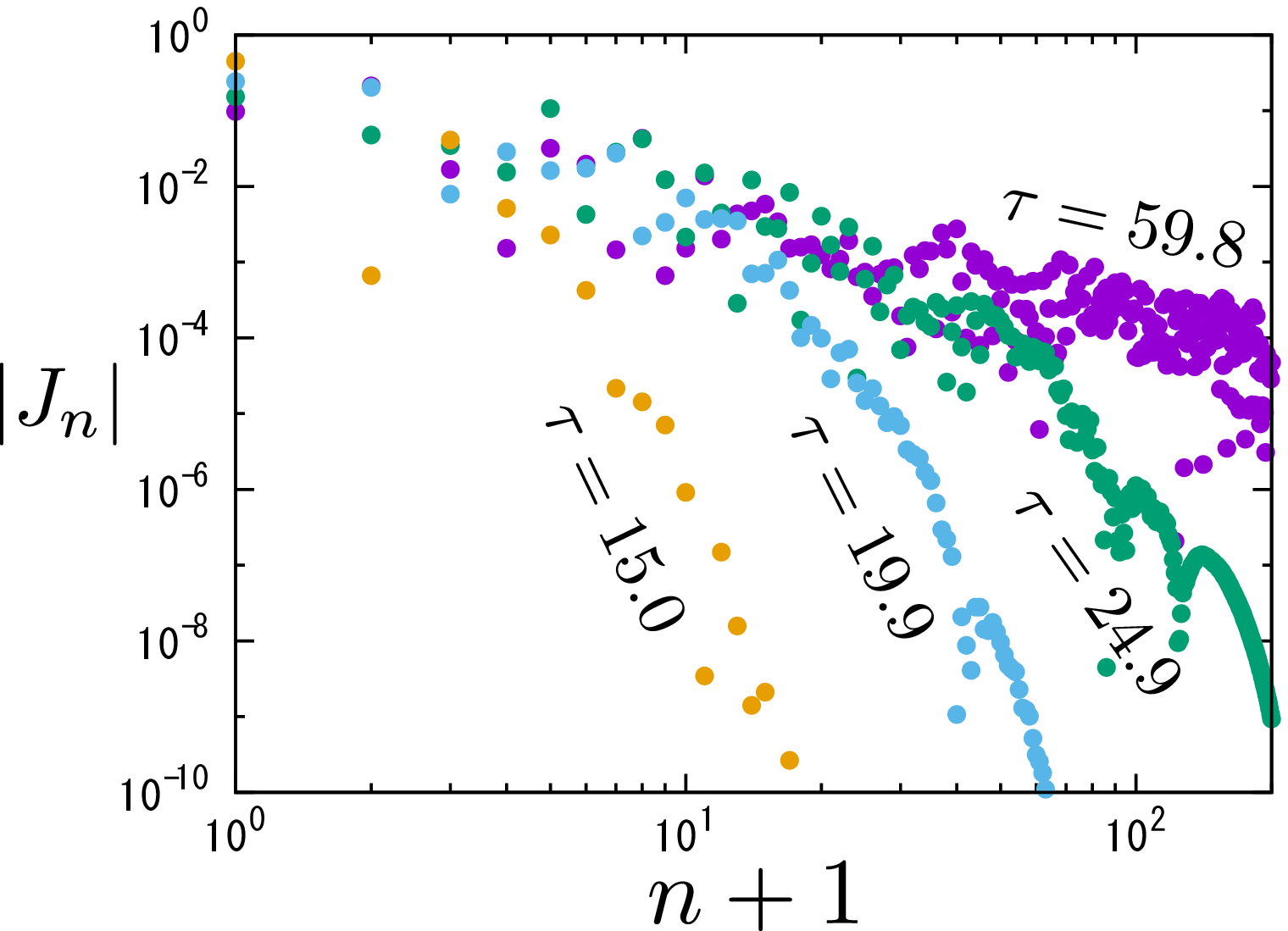}\label{Jspec1}
 }
 \subfigure[$E=1.11$, $J=0.5$]
  {\includegraphics[scale=0.45]{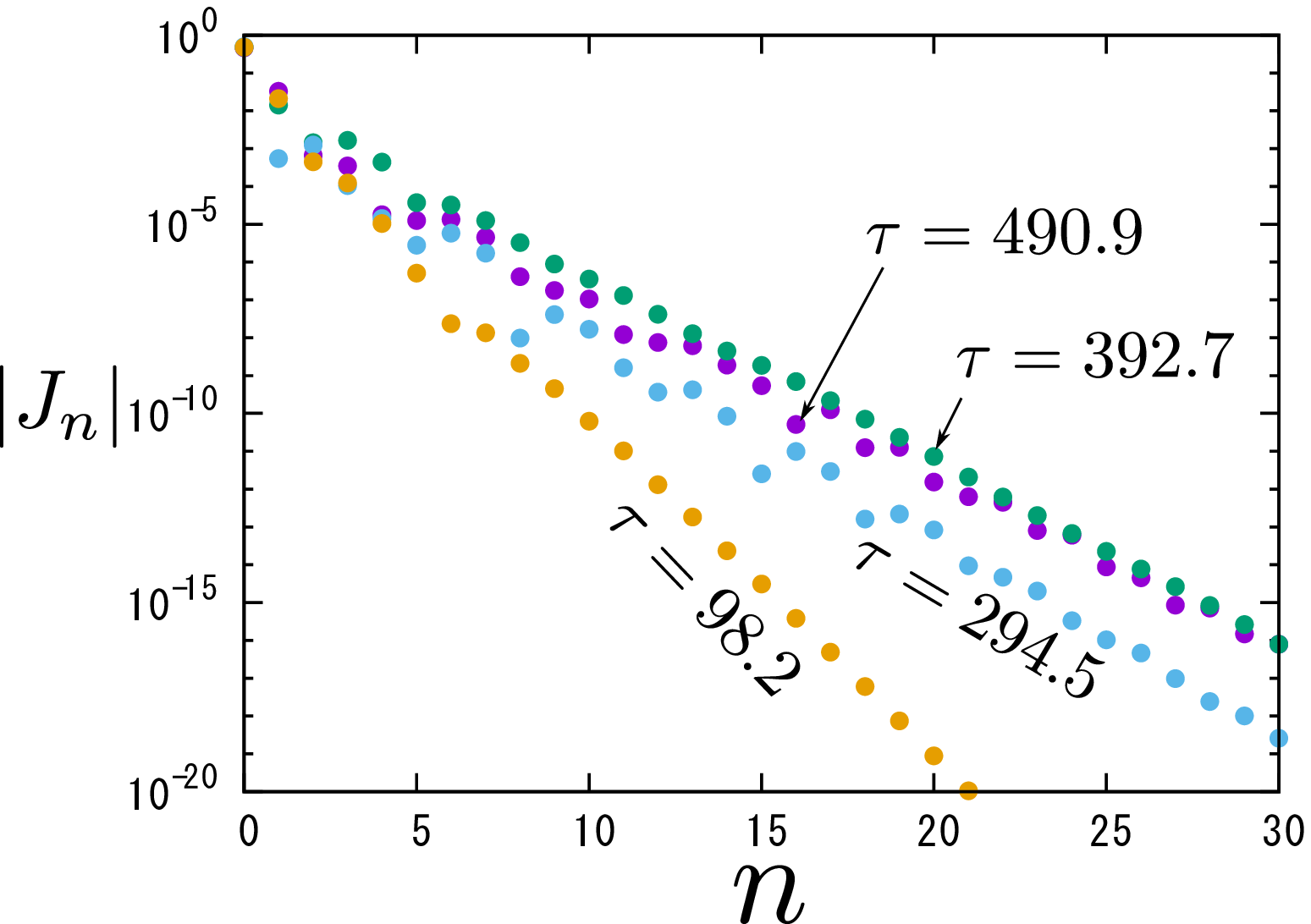}\label{Jspec2}
 }
  \caption{%
 Angular momentum spectrum for several values of $\tau$.
 The parameters for these figures are the same as those in Fig.~\ref{LandS}.
 The left and right panels are shown in log-log and semi-log scales, respectively.
 }
 \label{Jspec}
\end{figure}

In Fig.~\ref{Jspec}, the spectrum of the angular momenta is plotted for several values of $\tau$ 
for the same parameters used in Fig.~\ref{LandS}. Figure \ref{Jspec1} exhibits 
the angular momentum flow from lower to higher modes. This behaviour is consistent 
with the fact that the spin part dominates the total angular momentum 
in the turbulent string condensation. Eventually, the spectrum becomes power law. 
In Fig.~\ref{Jspec2}, the exponential spectrum is always indicated even in late times 
(Note here that the right figure is plotted at the semi-log scale). 

\begin{figure}
\begin{center}
\includegraphics[scale=0.5]{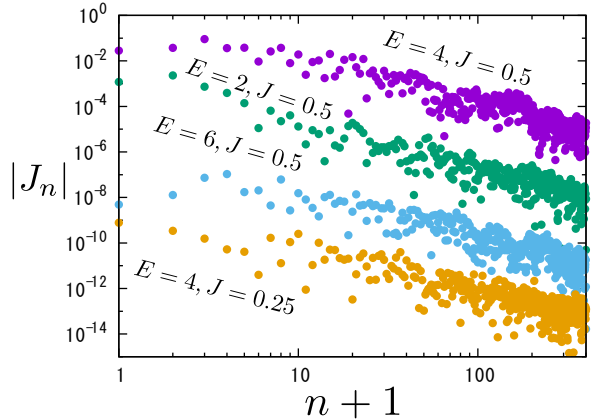}
\end{center}
 \caption{%
Angular momentum spectra at late time. 
We see the universal power law for the turbulent string condensation.
}
\label{univpow}
\end{figure}

Does the late time power law in the spectrum of the angular momenta exhibit a universal scaling?
To check universality, let us consider the following different initial data (i)-(iv):
\begin{enumerate}
\renewcommand{\labelenumi}{(\roman{enumi})}
\item $(E,J)=(4.0,0.5)$: $(\epsilon,r_0,\omega,\rho_0)=(0.02,0.4,0.4289,0.7939)$
\item $(E,J)=(2.0,0.5)$: $(\epsilon,r_0,\omega,\rho_0)=(0.02,0.4,0.4289,0.3522)$
\item $(E,J)=(6.0,0.5)$: $(\epsilon,r_0,\omega,\rho_0)=(0.02,0.4,0.4289,1.2140)$
\item $(E,J)=(4.0,0.25)$: $(\epsilon,r_0,\omega,\rho_0)=(0.02,0.4,0.2145,0.8144)$
\end{enumerate}
The initial data (i) is the same as the initial condition II used in Sec.~\ref{sec:string_lyapunov}.
We have checked that these four choices resulted in the turbulent string condensation, 
while the timescales to reach the power law scaling are different depending on the parameters.
In Fig.~\ref{univpow}, we show the angular momentum spectra at late times:
$\tau=99.7, 299.1, 39.9, 99.7$ for (i)-(iv), respectively. For visibility,
we multiplied $10^{-2}$, $10^{-6}$, $10^{-8}$ to the spectra of (ii)-(iv), respectively.
These plots seem to have a universal scaling. In fact, fitting the late time spectra 
with $\propto (n+1)^{-a}$ in $10\leq n+1 \leq 400$, we obtain $a=2.12, 1.94, 2.09, 1.85$ 
for (i)-(iv), respectively. These result may be consistent with a universal scaling of the spectra 
given by $|J_n|\propto n^{-2}$.

\subsection{Energy distribution}

\begin{figure}
  \centering
  \subfigure[Energy spectrum for $E=4.0$, $J=0.5$]
  {\includegraphics[scale=0.45]{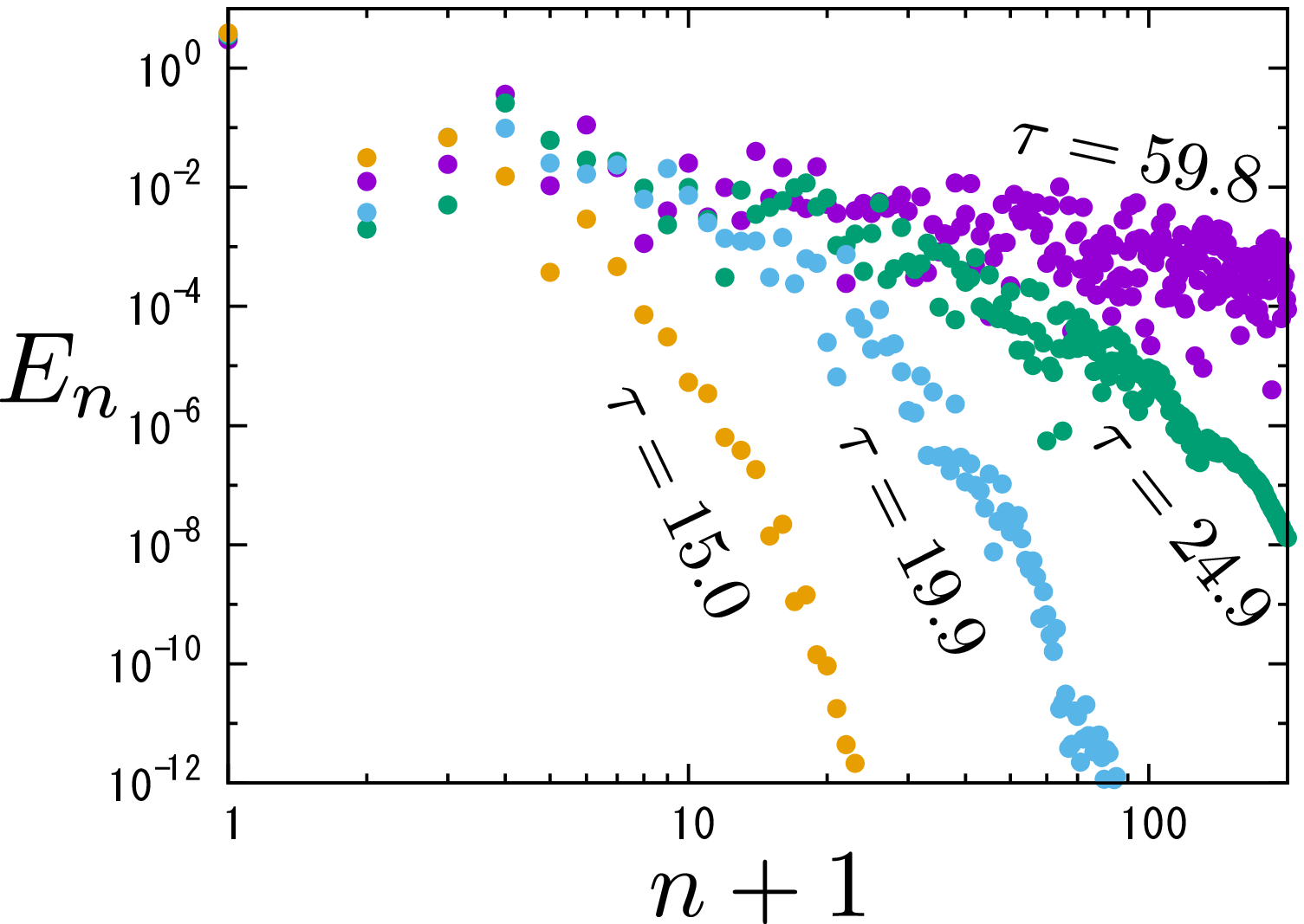}\label{Espec1}
 }
  \subfigure[Power law spectra at late time]
  {\includegraphics[scale=0.45]{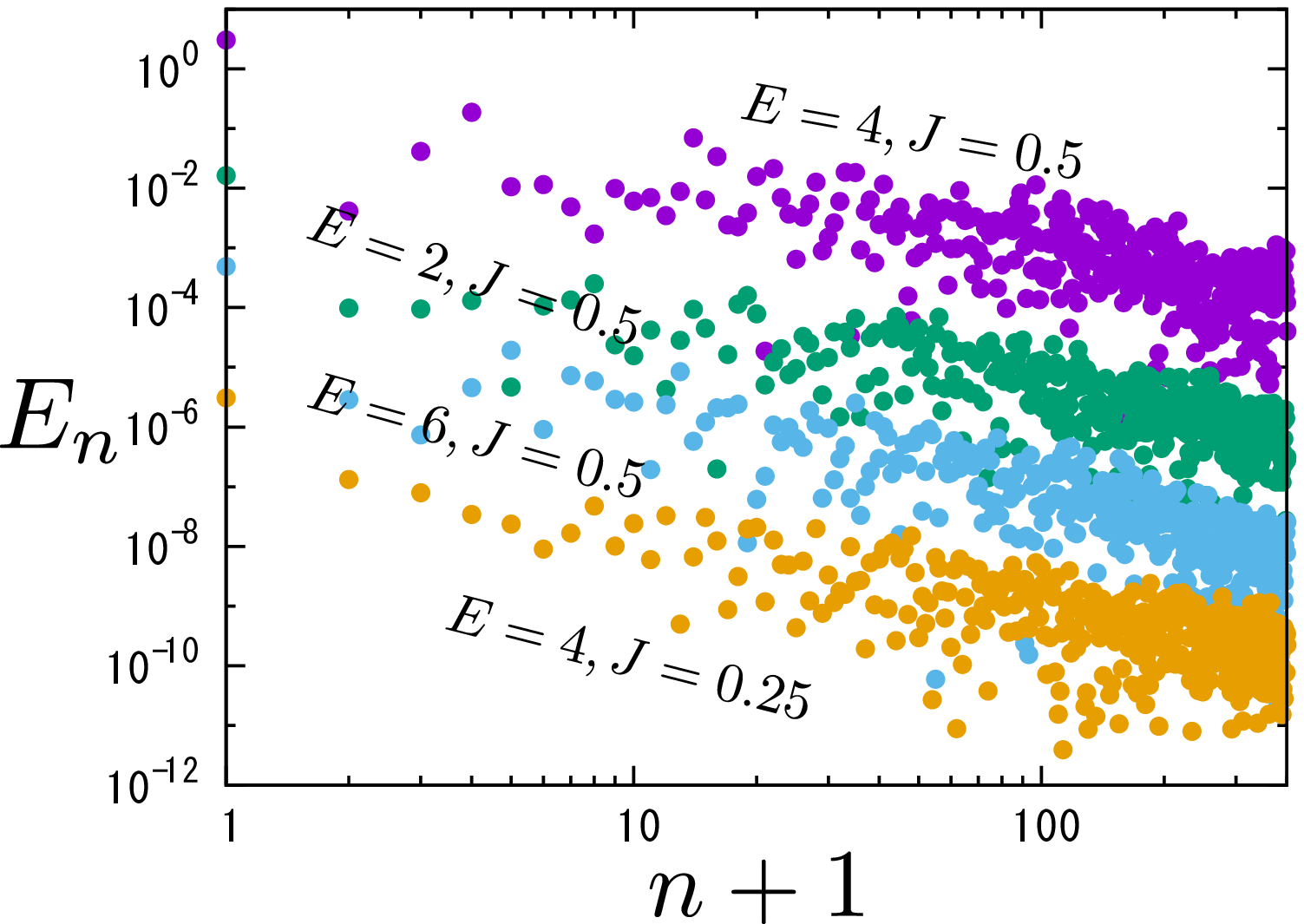}\label{Espec2}
 }
  \caption{%
 (a)~Energy spectrum at several time slices for $E=4.0$ and $J=0.5$.
 The spectrum is power law at late time.
 (b)~Energy spectra at late time for several parameters.
 The power does not depend on the parameters so much.
 }
 \label{Espec}
\end{figure}

To define the energy spectrum, let us consider the Fourier transformation of $\sqrt{p_t}$ as
\begin{equation}
 \sqrt{p_t(\tau,\sigma)}=\sum_{n=-\infty}^\infty c_n(\tau)e^{in\sigma}\ .
\end{equation}
Substituting this into the first equation of~(\ref{Ek}), we obtain
\begin{equation}
 E=\sum_{n=0}^\infty E_n\ ,\qquad
 E_0\equiv|c_0|^2\ ,\quad E_n\equiv|c_n|^2+|c_{-n}|^2\ .
\end{equation}
In Fig.~\ref{Espec1}, the energy spectra are plotted for several values of $\tau$ in the case of  
the initial condition (i) introduced in the previous subsection. Again, one can observe 
the energy flow from large to small scales, and eventually the spectrum obeys a power law.
In Fig.~\ref{Espec2}, the energy spectra for the conditions (i)-(iv) are shown at late times:
$\tau=99.7, 299.1, 39.9, 99.7$ for (i)-(iv), respectively. 
For visibility, the factors $10^{-2}$, $10^{-4}$, $10^{-6}$ have been multiplied 
to the spectra of (ii)-(iv), respectively. By sight, the power law exponents seem universal. 
Fitting the late time spectra with $\propto (n+1)^{-a}$ in $10\leq n+1 \leq 400$, we obtain 
$a=1.28$, $1.31$, $1.63$, $1.13$ for (i)-(iv), respectively.\footnote{
If we change the fitting region, the power is slightly changed: For the initial condition (i), 
the power is given by $1.22$ and $1.62$ for fitting regions 
$1\leq n+1 \leq 400$ and $100\leq n+1 \leq 400$, respectively.}
In the fits, the universality of the power is not so clear as in the case of the spectrum of 
the angular momenta, but it may be possible to say that the power does not depend 
on the parameters so much 
(Note that the energy of (ii) is three times larger than that of (iii)).

\section{Summary and discussion}
\label{sec:summary}

We have discussed chaotic and turbulent behaviors of classical strings in a 5D AdS soliton spacetime. 
First, we revisited classical chaos on this background with a cohomogeneity-1 string ansatz. 
We computed Poincar\'e sections and a Lyapunov spectrum and found chaos in this system. 
Then, we considered classical strings including the dependence on the spatial direction of 
the string world-sheet. In the chaotic parameter regime, we found that the string tends 
to stay around the tip of the AdS soliton with a completely jumbled profile. Because of the chaos,
trajectories of each string segment (expanded initially with a radius $\epsilon=0.02$) spread 
at the Lyapunov timescale, and the string extends in the target space at this timescale.
This can be indeed explained by considering the Fourier transform of the string 
along the spatial direction, where higher mode coefficients grow exponentially in time.
We also studied the time dependence of the orbital angular momentum 
and internal spin of the string. At the early stage of the time evolution, 
the total angular momentum is dominated by the orbital one that the initial configuration has.
However, it is transferred to the internal spin, and eventually the internal spin exceeds
the orbital angular momentum. We also studied the angular momentum and energy spectra
and found turbulent behaviour: At late times, the spectra realize
universal power-law scalings. The angular momentum transfer due to the turbulence is responsible 
for reaching the jumbled profile, and we referred this behaviour as the turbulent string condensation.

There are a lot of open questions. So far, we have considered classical solutions 
of strings whose shapes are not strictly protected by symmetries. 
These should correspond to certain composite operators according to the AdS/CFT 
dictionary and it should be possible to identify the correspondence. 
Unfortunately, we have not gotten a definite answer to this question yet. A naive candidate 
is a composite operator in which fields are aligned in a complex way and the complexity would be
closely related to the fractal structure of the classical chaos on the string-theory side. 
Also, this randomness implies dynamical information loss and it should be related to 
the production of the Kolmogorov-Sinai entropy at least in a chaotic parameter region, 
in which Pesin's equality holds and our Lyapunov exponents should be concerned with 
this direction. 

It would be interesting to study the cascading process of the turbulent strings in more detail. 
When the deformation parameter in this confined geometry vanishes, the original AdS$_5$ 
background is reproduced, which is known to be integrable. Hence it would be interesting 
to study and distinguish the long-time behaviours between the confined geometry and the AdS$_5$. 
This will lead to help understand more about non-integrable and integrable behaviours in the string dynamics.

We can also consider open strings in the AdS soliton space time.
The open string hanging from the AdS boundary corresponds
to a confined quark-antiquark pair in the dual gauge theory. 
Such an open string would show the turbulent behaviour once we take into account its time dependence.
It would be interesting to study how the turbulence affects the quark-antiquark pair in the confining phase.

Our results suggest that strings in non-integrable background geometries expand and form jumbled profiles.
This behaviour reminds us of the fuzzball conjecture:
Bound states of D-branes forming BPS black holes expand to a size that depends on their degeneracy \cite{Lunin:2001jy,Lunin:2002qf}.
In this picture, the black hole horizons are replaced by a fuzz of fluctuating strings or D-branes. 
It would be intriguing to consider a relation between the turbulent string condensation
and the fuzzball-like interpretation.\footnote{We thank Koji Hashimoto for pointing out this similarity.} 

We hope that our results on turbulent strings would shed light 
on the associated interpretation on the gauge-theory side.

\acknowledgments{
We are very grateful to Koji Hashimoto for fruitful comments.
The work of T.I.\ was supported by the Department of Energy, DOE award No.~DE-SC0008132.
The work of K.M.\ was supported by JSPS KAKENHI Grant Number 15K17658.
The work of K.Y.\ was supported by the Supporting Program for Interaction-based 
Initiative Team Studies (SPIRITS) from Kyoto University and by a JSPS Grant-in-Aid 
for Scientific Research (C) No.\,15K05051. This work was also supported in part 
by the JSPS Japan-Russia Research Cooperative Program 
and the JSPS Japan-Hungary Research Cooperative Program.  
}

\appendix

\section*{Appendix}

\section{Initial data construction}
\label{app:init}

In this appendix, we explain how to construct initial data satisfying 
the constraint equations~(\ref{CON}).
To give the data, we use the polar coordinates on the $(y_1,y_2)$-plane: 
$y_1=\rho \cos\phi$ and $y_2=\rho\sin\phi$. 
Then the constraints are rewritten as
\begin{align}
&-t_{,u}^2 +\rho_{,u}^2+\rho^2 \phi_{,u}^2
 + H(\bm{\chi}^2)\bm{\chi}_{,u}^2=0\ ,\label{CONnew1}\\
&-t_{,v}^2 +\rho_{,v}^2+\rho^2 \phi_{,v}^2
 + H(\bm{\chi}^2)\bm{\chi}_{,v}^2=0\ ,\label{CONnew2}
\end{align}
where $H\equiv G/F$.

\begin{figure}
\begin{center}
\includegraphics[scale=0.35]{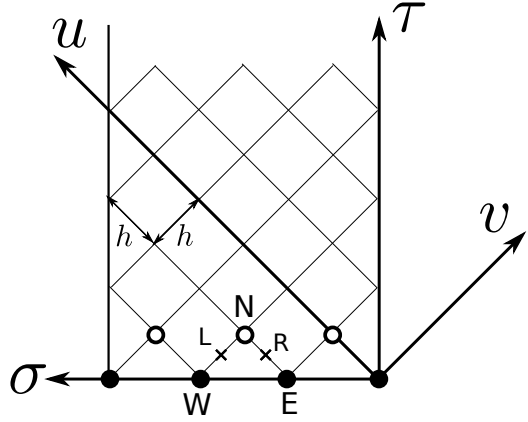}
\end{center}
 \caption{%
 String worldsheet with a uniform grid along the double null coordinates.
 Initial data are specified at black points $\bullet$ and white points $\circ$.
}
\label{worldsheet}
\end{figure}

In Fig.~\ref{worldsheet}, we show the numerical domain on the string worldsheet, where 
a uniform grid is taken with the spacing $h$ along the double null coordinates $u$ and $v$. 
Since the closed string is considered, the periodic boundary condition should be imposed:
$\sigma\sim \sigma+2\pi$. 
(The orthogonal coordinates $(\tau,\sigma)$ are introduced in Eq.~(\ref{tau_sigma}).)
The evolution and constraint equations are discretized on this lattice.

We specify initial data at $\tau=0$ and $\tau=h$, which are shown in black ($\bullet$) 
and white points ($\circ$) in the figure. 
At the black points ($\tau=0$), the initial data is given as in
Eqs.~(\ref{chiInit}) and (\ref{tyinit}).
At the white points ($\tau=h$), we rotate the string in the $(\chi_1,\chi_2)$-plane 
keeping the configuration in Eq.~(\ref{chiInit}) as
\begin{equation}
 (\chi_1+i\chi_2)|_{\tau=h}=e^{i\omega h}(\chi_1+i\chi_2)|_{\tau=0}\ ,
  \label{chi_tauh}
\end{equation}
where $\omega$ represents the initial angular velocity of the string
in the $\bm{\chi}$-plane (in terms of the worldsheet time coordinate $\tau$).
In the $\vec{y}$-plane, we consider the circular string with the same radius 
as that at $\tau=0$: $\rho|_{\tau=h}=\rho_0$.

Other variables $t|_{\tau=h}$ and $\phi|_{\tau=h}$
are determined by the constraint equations.
Now, we focus on points N, E, W, R, and L in Fig.~\ref{worldsheet}, where
R(L) are located in the middle of N and E(W).
Using the discretized data at N, E and W, 
we evaluate the constraint equations~(\ref{CONnew1}) and (\ref{CONnew2}) 
at R and L respectively:
\begin{align}
&-t_N^2+\rho_0^2 (\phi_N-\phi_E)^2 + H(\bm{\chi}_R^2)(\bm{\chi}_N-\bm{\chi}_E)^2=0\ ,
 \label{Con1disc}\\
&-t_N^2+\rho_0^2 (\phi_N-\phi_W)^2 + H(\bm{\chi}_L^2)(\bm{\chi}_N-\bm{\chi}_W)^2=0\ ,
\label{Con2disc}
\end{align}
where we have used second order central finite differentials for the derivatives, 
$t_E=t_W=0$, and $\rho_N=\rho_W=\rho_E=\rho_0$ and also defined
$\bm{\chi}_R=(\bm{\chi}_N+\bm{\chi}_E)/2$
and $\bm{\chi}_L=(\bm{\chi}_N+\bm{\chi}_W)/2$.
Note that $\bm{\chi}_{N,E,W}$ and $\phi_{E,W}$ are already given by
Eqs.~(\ref{chiInit}), (\ref{tyinit}) and (\ref{chi_tauh}). 
Only $t_N$ and $\phi_N$ are unknown in (\ref{Con1disc}) and (\ref{Con2disc}).
From these, we obtain 
\begin{equation}
\phi_N
 =\frac{1}{2}\left[
\phi_W+\phi_E
+\frac{H(\bm{\chi}_L^2)(\bm{\chi}_N-\bm{\chi}_W)^2
-H(\bm{\chi}_R^2)(\bm{\chi}_N-\bm{\chi}_E)^2}{\rho_0^2 (\phi_W-\phi_E)}\right]\ .
\end{equation}
Substituting the above equation into Eq.~(\ref{Con1disc}) leads to $t_N$.

\section{Error analysis}

In this section, let us estimate discretization numerical errors using constraint violation.
The constraint equations (\ref{CON}) can be rewritten as 
\begin{equation}
 \tilde{C}_1=-t_{,u}^2+\vec{y}_{,u}^{\,2}+H(\bm{\chi}^2) \bm{\chi}_{,u}^2=0\ ,\quad
 \tilde{C}_2=-t_{,v}^2+\vec{y}_{,v}^{\,2}+H(\bm{\chi}^2) \bm{\chi}_{,v}^2=0\ .
\end{equation}
We define a normalized constraint as
\begin{equation}
 C(\tau,\sigma)=\frac{|\tilde{C}_1|+|\tilde{C}_2|}{t_{,u}^2+\vec{y}_{,u}^{\,2}+H(\bm{\chi}^2) \bm{\chi}_{,u}^2 + t_{,v}^2+\vec{y}_{,v}^{\,2}+H(\bm{\chi}^2) \bm{\chi}_{,v}^2}\ .
\end{equation}
This is a function of two variables $\tau$ and $\sigma$. 
For convenience in evaluating the constraint violation,
the maximum value of $C(\tau,\sigma)$ on each $\tau$-slice is taken as
\begin{equation}
C_\textrm{max}(\tau)=\max_{0\leq \sigma < 2\pi} C(\tau,\sigma)\ .
\end{equation}
This is a function of a single variable $\tau$.

\begin{figure}
\begin{center}
\includegraphics[scale=0.5]{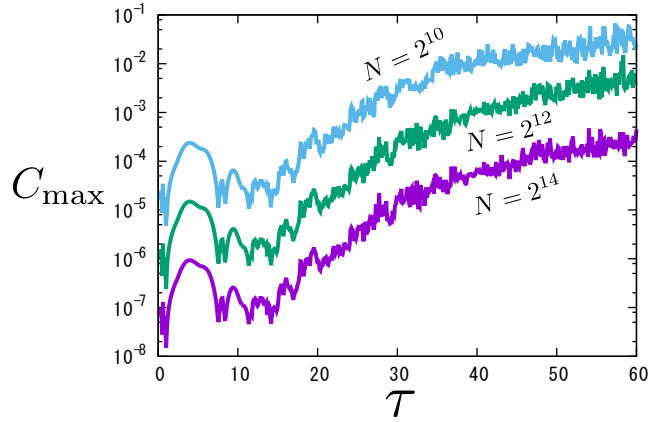}
\end{center}
\caption{%
Constraint violation at several resolutions, $N=2^{10}, 2^{12}, 2^{14}$.
The parameters are the same as those used in Fig.~\ref{profile}.
}
\label{Cmax}
\end{figure}

In Fig.~\ref{Cmax}, $C_\textrm{max}(\tau)$ is plotted at several resolutions 
$N=2^{10}, 2^{12}, 2^{14}$ with $(\epsilon,r_0,\omega,\rho_0)=(0.02,0.4,0.4289,0.7939)$ 
which are the same parameters as those used in Fig.~\ref{profile}.
Here, $N$ is the number of grid points along the $\sigma$-direction
(The numbers of black and white points in Fig.~\ref{worldsheet} are $N+1$ and $N$, respectively).
One can find that the constraint violation is certainly small ($C_\textrm{max}<10^{-3}$ 
for $N=2^{14}$ and $\tau\leq 60$). We use a second-order discretization scheme, 
and the scaling $C_\textrm{max}\propto 1/N^2$ is consistent with this.
In the paper, we work with $N=2^{14}$ as a typical value of $N$.

\begin{thebibliography}{99}

\bibitem{M}  
 J.~M.~Maldacena, 
 ``The large N limit of superconformal field theories and supergravity,''
 Adv.\ Theor.\ Math.\ Phys.\  {\bf 2} (1998) 231
 [Int.\ J.\ Theor.\ Phys.\  {\bf 38} (1999) 1113].  
[arXiv:hep-th/9711200].

\bibitem{GKP}
S.~S.~Gubser, I.~R.~Klebanov and A.~M.~Polyakov,
``Gauge theory correlators from non-critical string theory,''
Phys.\ Lett.\ B {\bf 428} (1998) 105 [arXiv:hep-th/9802109]. 

\bibitem{W}
E.~Witten, 
``Anti-de Sitter space and holography,''
Adv.\ Theor.\ Math.\ Phys.\  {\bf 2} (1998) 253 [arXiv:hep-th/9802150].

\bibitem{review}
 N.~Beisert {\it et al.},
 ``Review of AdS/CFT Integrability: An Overview,'' 
 Lett.\ Math.\ Phys.\ {\bf 99} (2012) 3 [arXiv:1012.3982 [hep-th]]. 

\bibitem{BPR}
I.~Bena, J.~Polchinski and R.~Roiban,
``Hidden symmetries of the AdS$_5\times$S$^5$ superstring,''
Phys.\ Rev.\ D {\bf 69} (2004) 046002 [hep-th/0305116].


\bibitem{Horowitz:1998ha} 
  G.~T.~Horowitz and R.~C.~Myers,
  ``The AdS / CFT correspondence and a new positive energy conjecture for general relativity,''
  Phys.\ Rev.\ D {\bf 59} (1998) 026005 
  [hep-th/9808079].
 
\bibitem{Basu:2011dg}
P.~Basu, D.~Das and A.~Ghosh,
  ``Integrability Lost,''
  Phys.\ Lett.\ B {\bf 699} (2011) 388
  [arXiv:1103.4101 [hep-th]].  

\bibitem{Candelas}
 P.~Candelas and X.~de la Ossa, 
 ``Comments on Conifolds,'' 
 Nucl.\ Phys.\ B {\bf 342} (1990) 246. 

\bibitem{KW}
I.~R.~Klebanov and E.~Witten,
``Superconformal field theory on three-branes at a Calabi-Yau singularity,''
Nucl.\ Phys.\ B {\bf 536} (1998) 199 [hep-th/9807080].
     
\bibitem{CMY} 
P.~M.~Crichigno, T.~Matsumoto and K.~Yoshida, 
``Deformations of $T^{1,1}$ as Yang-Baxter sigma models,'' 
JHEP {\bf 1412} (2014) 085 
[arXiv:1406.2249 [hep-th]]; 
``Towards the gravity/CYBE correspondence beyond integrability 
-- Yang-Baxter deformations of $T^{1,1}$,'' 
J.\ Phys.\ Conf.\ Ser.\  {\bf 670} (2016) no.1,  012019 
[arXiv:1510.00835 [hep-th]]. \\ 
I.~Kawaguchi, T.~Matsumoto and K.~Yoshida, 
``Jordanian deformations of the AdS$_5 \times$S$^5$ superstring,'' 
JHEP {\bf 1404} (2014) 153 [arXiv:1401.4855 [hep-th]].  
  
\bibitem{T11}
P.~Basu and L.~A.~Pando Zayas,
  ``Chaos Rules out Integrability of Strings in AdS$_5 \times T^{1,1}$,''  
Phys.\ Lett.\ B {\bf 700} (2011) 243 [arXiv:1103.4107 [hep-th]]; 
  ``Analytic Non-integrability in String Theory,''
  Phys.\ Rev.\ D {\bf 84} (2011) 046006
  [arXiv:1105.2540 [hep-th]].   
  
\bibitem{T11-ppwave} 
  Y.~Asano, D.~Kawai, H.~Kyono and K.~Yoshida,
  ``Chaotic strings in a near Penrose limit of AdS$_{5} \times T^{1,1}$,''
  JHEP {\bf 1508} (2015) 060 
  [arXiv:1505.07583 [hep-th]].

  
\bibitem{BH}
 L.~A.~Pando Zayas and C.~A.~Terrero-Escalante,
  ``Chaos in the Gauge / Gravity Correspondence,''
  JHEP {\bf 1009} (2010) 094
  [arXiv:1007.0277 [hep-th]].\\ 
P.~Basu, P.~Chaturvedi and P.~Samantray, 
``Chaotic dynamics of strings in charged black hole backgrounds,'' 
arXiv:1607.04466. 

 \bibitem{WQCD} 
 P.~Basu, D.~Das, A.~Ghosh and L.~A.~Pando Zayas,
  ``Chaos around Holographic Regge Trajectories,''
  JHEP {\bf 1205} (2012) 077
  [arXiv:1201.5634 [hep-th]]. \\ 
L.~A.~Pando Zayas and D.~Reichmann, 
``A String Theory Explanation for Quantum Chaos in the Hadronic Spectrum,'' 
JHEP {\bf 1304} (2013) 083 
[arXiv:1209.5902 [hep-th]]. 

\bibitem{D-brane}
A.~Stepanchuk and A.~A.~Tseytlin,
  ``On (non)integrability of classical strings in p-brane backgrounds,''
  J.\ Phys.\ A {\bf 46} (2013) 125401
  [arXiv:1211.3727 [hep-th]]. \\ 
   Y.~Chervonyi and O.~Lunin,
  ``(Non)-Integrability of Geodesics in D-brane Backgrounds,''
  JHEP {\bf 1402} (2014) 061
  [arXiv:1311.1521 [hep-th]].
  
\bibitem{complex-beta}  
D.~Giataganas, L.~A.~Pando Zayas and K.~Zoubos,
  ``On Marginal Deformations and Non-Integrability,''
  JHEP {\bf 1401} (2014) 129
  [arXiv:1311.3241 [hep-th]].  

\bibitem{NR}
  D.~Giataganas and K.~Sfetsos,
  ``Non-integrability in non-relativistic theories,''
  JHEP {\bf 1406} (2014) 018
  [arXiv:1403.2703 [hep-th]]. \\ 
 X.~Bai, J.~Chen, B.~H.~Lee and T.~Moon,
  ``Chaos in Lifshitz Spacetimes,'' 
   J.\ Korean Phys.\ Soc.\  {\bf 68} no.5 (2016)  639 
  [arXiv:1406.5816 [hep-th]].
  
\bibitem{gamma} 
  K.~L.~Panigrahi and M.~Samal,
  ``Chaos in classical string dynamics in $\hat{\gamma}$ deformed AdS$_5 \times T^{1,1}$,''
  arXiv:1605.05638 [hep-th].  
  
 \bibitem{Melnikov1}
  V.~K.~Melnikov,
  ``On the stability of the center for time periodic perturbations,''
  Trans.\ Moscow Math.\ Soc.\ {\bf 12} (1963) 1.

\bibitem{Melnikov2} 
  P.~J.~Holmes and J.~E.~Marsden,
  ``Horseshoes in Perturbations of Hamiltonian Systems with Two Degrees of Freedom,''
  Commun.\ Math.\ Phys.\ {\bf 82} (1982) 523.

 \bibitem{AKY}
 Y.~Asano, H.~Kyono and K.~Yoshida,
  ``Melnikov's method in String Theory,''
 JHEP {\bf 1609} (2016) 103
  [arXiv:1607.07302 [hep-th]]. 
  
\bibitem{HY}
S.~A.~Hartnoll and K.~Yoshida,
  ``Families of IIB duals for nonrelativistic CFTs,''
  JHEP {\bf 0812} (2008) 071
  [arXiv:0810.0298 [hep-th]]. 
 
  
\bibitem{Koike:2008fs}
  T.~Koike, H.~Kozaki and H.~Ishihara,
 ``Strings in five-dimensional anti-de Sitter space with a symmetry,''
  Phys.\ Rev.\ D {\bf 77} (2008) 125003
  [arXiv:0804.0084 [gr-qc]].




\bibitem{BFSS}
  T.~Banks, W.~Fischler, S.~H.~Shenker and L.~Susskind,
  ``M theory as a matrix model: A Conjecture,''
  Phys.\ Rev.\ D {\bf 55} (1997) 5112 [hep-th/9610043]. 
  
\bibitem{BMN}
  D.~E.~Berenstein, J.~M.~Maldacena and H.~S.~Nastase,
 ``Strings in flat space and pp waves from N=4 super Yang-Mills,''
  JHEP {\bf 0204} (2002) 013 [hep-th/0202021].     
  
\bibitem{chaos-BFSS}
  I.~Y.~Aref'eva, P.~B.~Medvedev, O.~A.~Rytchkov and I.~V.~Volovich,
  ``Chaos in M(atrix) theory,''
  Chaos Solitons Fractals {\bf 10} (1999) 213
  [hep-th/9710032].
 
\bibitem{Berenstein} 
 C.~T.~Asplund, D.~Berenstein and D.~Trancanelli,
  ``Evidence for fast thermalization in the plane-wave matrix model,''
  Phys.\ Rev.\ Lett.\  {\bf 107} (2011) 171602
  [arXiv:1104.5469 [hep-th]]; 
 C.~T.~Asplund, D.~Berenstein and E.~Dzienkowski,
  ``Large N classical dynamics of holographic matrix models,''
  Phys.\ Rev.\ D {\bf 87} (2013) 8,  084044
  [arXiv:1211.3425 [hep-th]]. 
    
 \bibitem{chaos-BMN} 
    Y.~Asano, D.~Kawai and K.~Yoshida,
  ``Chaos in the BMN matrix model,'' 
   JHEP {\bf 1506} (2015) 191
  [arXiv:1503.04594 [hep-th]]. 
  
\bibitem{YM}
G.~Z.~Baseyan, S.~G.~Matinyan and G.~K.~Savvidi, 
``Nonlinear plane waves in the massless Yang-Mills theory,'' 
JETP Lett.\ {\bf 29} (1979) 587.  \\ 
B.~V.~Chirikov and D.~L.~Shepelyanskii, 
``Stochastic oscillations of classical Yang-Mills fields,'' 
JETP Lett.\ {\bf 34} (1981) 163.    

\bibitem{deformed-YM}  
S.~G.~Matinyan, G.~K.~Savvidi and N.~G.~Ter-Arutyunyan-Savvidi, 
``Stochasticity of classical Yang-Mills mechanics and its elimination by using the Higgs mechanism,'' 
JETP Lett.\ {\bf 34} (1981) 590. 




\bibitem{HMY}
  K.~Hashimoto, K.~Murata and K.~Yoshida,
  ``Chaos of chiral condensate,'' \\
  arXiv:1605.08124 [hep-th].  
  
 
\bibitem{Ishii:2015wua} 
  T.~Ishii and K.~Murata,
  ``Turbulent strings in AdS/CFT,''
  JHEP {\bf 1506} (2015) 086 
  [arXiv:1504.02190 [hep-th]].
  
\bibitem{Hashimoto:2014yza} 
  K.~Hashimoto, S.~Kinoshita, K.~Murata and T.~Oka,
  ``Electric Field Quench in AdS/CFT,''
  JHEP {\bf 1409} (2014) 126 
  [arXiv:1407.0798 [hep-th]].


\bibitem{Hashimoto:2014xta} 
  K.~Hashimoto, S.~Kinoshita, K.~Murata and T.~Oka,
  ``Turbulent meson condensation in quark deconfinement,''
  Phys.\ Lett.\ B {\bf 746} (2015) 311 
  [arXiv:1408.6293 [hep-th]].


\bibitem{Hashimoto:2014dda} 
  K.~Hashimoto, S.~Kinoshita, K.~Murata and T.~Oka,
  ``Meson turbulence at quark deconfinement from AdS/CFT,''
  Nucl.\ Phys.\ B {\bf 896} (2015) 738 
  [arXiv:1412.4964 [hep-th]].





\bibitem{Ko}
A.~N.~Kolmogorov, ``The conservation of conditionally periodic 
motion with a small variation in the Hamiltonian,'' 
Dokl.\ Akad.\ Nauk SSSR {\bf 98} (1954) 527. 

\bibitem{Ar}
V.~I.~Arnold, ``Small denominators and problems of stability of motion 
in classical and celestial mechanics,''
Uspekhi Mat.\ Nauk, Russian Math.\ {\bf 18} No.\ 6 (1963) 91; 
Russ.\ Math.\ Surv.\ {\bf 18} (1963) 9.  

\bibitem{Mo}
J.~Moser, ``On invariant curves of area-preserving mappings of an annulus,''
Nachr.\ Akad.\ Wiss.\ G\"ottingen Math.-Phys.\ Kl.\ II (1962) 1. 

\bibitem{Shimada}
I.~Shimada and T.~Nagashima, 
``A Numerical Approach to Ergodic Problem of Dissipative Dynamical Systems,'' 
Prog.\ Theor.\ Phys.\ {\bf 61} (1979) 1605.


	
\bibitem{Ishii:2014paa}
  T.~Ishii, S.~Kinoshita, K.~Murata and N.~Tanahashi,
  ``Dynamical Meson Melting in Holography,''
  JHEP {\bf 1404} (2014) 099
  [arXiv:1401.5106 [hep-th]].

\bibitem{Ishii:2015qmj} 
  T.~Ishii and K.~Murata,
  ``Dynamical AdS strings across horizons,''
  JHEP {\bf 1603} (2016) 035 
  [arXiv:1512.08574 [hep-th]].

\bibitem{Lunin:2001jy} 
  O.~Lunin and S.~D.~Mathur,
  ``AdS / CFT duality and the black hole information paradox,''
  Nucl.\ Phys.\ B {\bf 623}, 342 (2002)
  [hep-th/0109154].



\bibitem{Lunin:2002qf} 
  O.~Lunin and S.~D.~Mathur,
  ``Statistical interpretation of Bekenstein entropy for systems with a stretched horizon,''
  Phys.\ Rev.\ Lett.\  {\bf 88}, 211303 (2002)
  [hep-th/0202072].

	



\end{thebibliography}
\end{document}